\documentclass{JHEP3}

\usepackage{graphicx}
\usepackage{amssymb}
\usepackage{amsmath}
\usepackage{cite}
\usepackage{afterpage}

\newcommand{\be}{\begin{equation}}
\newcommand{\en}{\end{equation}}
\newcommand{\bea}{\begin{eqnarray}}
\newcommand{\ena}{\end{eqnarray}}

\newcommand{\mean}[1]{\left\langle #1 \right\rangle}
\newcommand{\mpl}{m_{_\mathrm{Pl}}}

\newcommand{\ie}{\textsl{i.e.}~}

\title{Dark Energy and the MSSM}

\author{Philippe Brax \thanks{Associate Researcher at Institut
d'Astrophysique de Paris, UMR 7095-CNRS, Universit\'e Pierre et Marie
Curie, 98bis boulevard Arago, 75014 Paris, France} \\ Service de
Physique Th\'eorique, CEA-Saclay, Gif/Yvette cedex, France F-91191 \\
E-mail: \email{brax@spht.saclay.cea.fr}}

\author{J\'er\^ome Martin \\ Institut d'Astrophysique de Paris, UMR
7095-CNRS, Universit\'e Pierre et Marie Curie, 98bis boulevard Arago,
75014 Paris, France \\ E-mail: \email{jmartin@iap.fr}} \date{\today}

\abstract{We consider the coupling of quintessence to observable
matter in supergravity and study the dynamics of both supersymmetry
breaking and quintessence in this context. We investigate how the
quintessence potential is modified by supersymmetry breaking and
analyse the structure of the soft supersymmetry breaking terms. We pay
attention to their dependence on the quintessence field and to the
electroweak symmetry breaking, \ie the pattern of fermion masses at
low energy within the Minimal Supersymmetric Standard Model (MSSM)
coupled to quintessence. In particular, we compute explicitly how the
fermion masses generated through the Higgs mechanism depend on the
quintessence field for a general model of quintessence. Fifth force
and equivalence principle violations are potentially present as the
vacuum expectation values of the Higgs bosons become quintessence
field dependent. We emphasize that equivalence principle violations
are a generic consequence of the fact that, in the MSSM, the fermions
couple differently to the two Higgs doublets. Finally, we also discuss
how the scaling of the cold dark and baryonic matter energy density is
modified and comment on the possible variation of the gauge coupling
constants, among which is the fine structure constant, and of the
proton-electron mass ratio.}

\begin{document}

\section{Introduction}

Cosmological observations combining the large scale structures of the
universe~\cite{LSS}, the Hubble diagram of type Ia
supernovae~\cite{IA} and the anisotropies of the cosmic microwave
background~\cite{CMB} all lead to the existence of an accelerated
phase of the expansion of the universe. Such an acceleration is
interpreted within Einsteinian relativity (see also Ref.~\cite{MSU})
as being driven by a vacuum energy of minute magnitude, some hundred
and twenty orders below the Planck scale. Most scenarios suggest that
such a very low value for the cosmological constant is hardly
compatible with known physics such as the electroweak phase transition
and the large radiative correction which are generated there. This has
prompted the possibility that the acceleration of the universe
expansion could have an extra dimensional origin. It could result from
a self-tuning mechanism where the energy density on a brane--world
curves a fifth dimension leaving an (almost) flat
brane~\cite{kachru}. It could also be that our universe undergoes a
phase of self-acceleration such as the ones existing in brane induced
gravity models~\cite{deffayet, Lalak,Koyama}. Another possibility is
that the vacuum energy has an anthropic origin coming from the
landscape of string vacua~\cite{Susskind}.

\par

In the following we will use a more traditional approach and utilize
four-dimensional field theory to tackle the problem of the
acceleration of the universe expansion. More precisely we will study
quintessence models~\cite{RP,quint,PB,BM1,cope}. As usual we consider
that the cosmological constant problem is in fact two--pronged. The
cosmological constant {\it per se} has to do with the exact
cancellation of large energy densities coming from Quantum Chromo
Dynamics (QCD), the electroweak phase transition, Grand Unified Theory
(GUT) scale physics and other high energy phenomena. Quintessence
models have nothing to say about this major problem. On the contrary
we concentrate on modeling the small vacuum energy responsible for the
acceleration. To do so, we assume that scalar field energy densities
are responsible for it. We also require that quintessence models
possess attractors~\cite{RP}, \ie long time solutions implying that no
sensitivity to initial conditions exists. Within such a class of
models, exemplified by the Ratra--Peebles model, the quintessence
field tends to have a large value now, very close to the Planck
scale. As a consequence we embed quintessence models within
supergravity (SUGRA)~\cite{BM1,BM2,BMR1,BMR2}, the best field
theoretic candidate capturing physics close to the Planck scale.

\par

In this paper, we are particularly interested in the coupling between
the quintessence field and the observable sector of particle physics
such as the Minimal Supersymmetric Standard Model (MSSM) or the mSUGRA
model~\cite{Nilles}. Such a model contains soft supersymmetry breaking
terms whose origin is best described using spontaneously broken
supersymmetry originating in a supersymmetry breaking sector. As
gravitational experiments give stringent bounds on fifth forces and
equivalence principle violations~\cite{GR} mediated by light scalar
fields such as the quintessence field, we impose that the quintessence
field lives in a separate sector from the supersymmetry breaking and
observable sectors. Hence we consider models with three
sectors. Nevertheless gravitational couplings between these sectors
are present and we study their consequences on the pattern of
supersymmetry breaking, the generation of soft terms~\cite{Brignole}
and the electroweak symmetry breaking. The soft terms and the fermion
masses after electroweak symmetry breaking~\cite{Savoy} all become
dependent on the quintessence field, \ie we have Yukawa-like
interaction of the form
\begin{equation}
\label{yukcoupling}
m^2\left(\frac{Q}{\mpl}\right)\bar{\Psi}\Psi \, ,
\end{equation}
where $\Psi $ represents a fermionic field and $Q$ is the quintessence
field. The main purpose of the paper is to calculate the function
$m\left(Q/\mpl\right)$ for a general model of quintessence. We also
study how the quintessence potential itself is modified by the
supersymmetry breaking and give the general expression of its new
shape. In particular, the value of the quintessence field now can be
drastically affected by the presence of the supersymmetry breaking
sector and can lead to an extremely low value of $Q$ compared to the
Planck scale~\cite{BMcosmo}.

\par

An interaction of the form~(\ref{yukcoupling}) is such that the model
reduces to a scalar--tensor--like theory~\cite{Brans} where matter
couples both to gravity and the quintessence field. In particular, we
find that matter couples to a different metric depending on which of
the two MSSM Higgs fields gives its mass to a particle.  This may lead
to large gravitational problems such as strong violations of the weak
equivalence principle~\cite{damour}, the presence of a fifth force or
the variation of the proton--electron mass ratio. Another consequence
of the interaction~(\ref{yukcoupling}) is that it implies the presence
of an interaction between dark energy and dark
matter~\cite{amendola,FP}. As a consequence, the scaling of the cold
dark and baryonic energy densities is modified and we briefly mention
the observational consequences of this fact.

\par

The paper is arranged as follows. In the next section we discuss the
general coupling between quintessence, supersymmetry breaking and
observable matter and calculate the resulting soft terms. In
section~\ref{ESB}, we apply these results to the electroweak symmetry
breaking. In section~\ref{Discussion}, we analyze different scenarios
and discuss the various physical consequences of the coupling between
the quintessence field and the fields of the standard model of
particles physics. The discussion is kept as general as possible while
concrete examples and applications of the calculations developed in
this article are studied in Ref.~\cite{BMcosmo}. Finally, in
Section~\ref{Conclusion}, we present our conclusions.

\section{Coupling Quintessence to Matter}
\label{Coupling Quintessence to Matter}

\subsection{General Setting}

In the following we will deal with the coupling of a quintessence
sector to an observable sector and a supersymmetry breaking
sector. The supersymmetry breaking sector is assumed to be only
gravitationally coupled to the observable sector as is standard in
SUGRA~\cite{Nilles}. Similarly we assume that the quintessence sector
is separated from the observable sector and couples to ordinary matter
only gravitationally. This is to prevent strong direct couplings
between matter and quintessence which may lead to the existence of a
fifth force or violations of the equivalence principle. In some sense,
this is also the minimal and most conservative assumption since, in
this case, the coupling between matter and quintessence sectors is
completely fixed by SUGRA and no other physical input is
necessary. Otherwise, it would have been necessary to choose an
explicit coupling and it is unclear which form it should take. Hence,
we assume that there are three different sectors in the theory: the
observable, hidden and quintessence sectors. As a consequence, the
K\"ahler and super potentials are given by the following expressions
\begin{eqnarray}
K &=& K_{\rm quint}+K_{\rm hid}+K_{\rm obs}\, , \quad W = W_{\rm
quint}+W_{\rm hid}+W_{\rm obs}\, .
\end{eqnarray}
Let us notice that the above choice, although very natural, is
probably not unique. For instance, another possibility inspired by the
physics of branes would be to consider the sequestered form~\cite{RS}
where one adds the exponential of the K\"ahler potentials, \ie
$g=-3/\kappa +g_{\rm quint}+g_{\rm hid}+g_{\rm obs}$ with $g\equiv
-(3/\kappa) {\rm e}^{-\kappa K/3}$, $\kappa $ being defined by $\kappa
\equiv 8\pi /\mpl^2$ rather than the K\"ahler potentials
themselves. Therefore, in this type of models, $K$ is given
by~\cite{RS}: $K=-(3/\kappa )\ln (1-\kappa g_{\rm quint}/3 -\kappa
g_{\rm hid}/3 -\kappa g_{\rm obs}/3)$. The analysis of the
phenomenology of these sequestered models is left for future work.

\par

Notice that we restrict our considerations to $N=1$ SUGRA in its usual
formulation limited to second order in the derivatives. Higher order
derivatives might give extra contributions such as $R^4$ where $R$ is
the Ricci scalar, see Ref.~\cite{GV}. However, these terms are Planck
mass suppressed and since one considers an effective theory valid up
to a cutoff small in comparison to the Planck mass (typically the GUT
scale), one expects these terms to be small. Otherwise, these
corrections could lead to K-essence models that have been studied in
Ref.~\cite{Mu} but are different from what is investigated in this
paper.

\par

In this article, the quintessence sector is left unspecified and no
assumption will be made about $K_{\rm quint}$ and $W_{\rm quint}$. In
this ``dark energy'' sector, we collectively denote the fields by
$d_{\alpha }$ (if it contains more than just one field) among which is
of course the quintessence field itself, $Q$. We denote the fields in
the hidden sector by $z_i$. For simplicity, we take a flat K\"ahler
potential and we do not specify the superpotential for the moment,
\begin{equation}
K_{\rm hid}=\sum _i z_iz_i^{\dagger }+\cdots \, ,\quad W_{\rm
  hid}=W_{\rm hid}(z_i)\, .
\end{equation}
Finally, the fields in the matter sector are written $\phi _a$.  This
sector is supposed to contain all the (super) fields that are
observable. As a consequence, we take this sector to be the MSSM or
the mSUGRA model~\cite{Nilles}, namely
\begin{equation}
\label{wobs}
K_{\rm obs}=\sum _a\phi _a\phi _a^{\dagger }+\cdots \, ,\quad W_{\rm
obs}=\frac13 \sum _{abc}\lambda _{abc}\phi_a \phi_b \phi_c +\frac12
\sum _{ab} \mu _{ab}\phi _a\phi _b+\cdots \, ,
\end{equation}
with a supersymmetric mass matrix $\mu_{ab}$ and Yukawa couplings
$\lambda_{abc}$. The dots denote possible extra terms suppressed by
the cutoff scale of the theory that will not be considered here. This
is compatible with the superpotential of the mSUGRA model. In the
following, when we treat the electroweak transition, we will be even
more specific about the form of the superpotential.

\par

In order to completely specify the observable sector, it is also
necessary to choose the supergravity gauge coupling functions
$f_{_{G}}$ appearing in the action as $\int {\rm d}^2\theta f_{_{G}}
W_G^2$ where $ W_{_{G}}$ is the superfield strength of the gauge group
$G$. If the function $f_{_{G}}$ depends on the quintessence field,
this leads to variations of the gauge coupling constants such as the
fine structure constant. In this article, we have chosen to keep the
more general dependence and assume that all the $f_{_{G}}$'s to be
$z_i$ and $d_{\alpha }$--dependent (hence, a priori, $Q$-dependent).

\par

Then, inserting the K\"ahler and the super potentials into the
expression of the scalar potential (the $F$--term), one gets
\begin{eqnarray}
\label{pottotal}
V &=& {\rm e}^{\kappa K}\left(K^{-1}\right)^{d_{\alpha
  }^{\dagger}d_{\beta }} \left(\kappa W\frac{{\partial }K_{\rm
  quint}}{\partial d_{\beta }} +\frac{{\partial }W_{\rm
  quint}}{\partial d_{\beta }}\right) \left(\kappa W^{\dagger
  }\frac{{\partial }K_{\rm quint}}{\partial d_{\alpha }^{\dagger }}
  +\frac{{\partial }W_{\rm quint}^{\dagger }}{\partial d_{\alpha
  }^{\dagger}}\right)\nonumber \\ & & + {\rm e}^{\kappa
  K}\left(K^{-1}\right)^{z_i^{\dagger}z_j} \left(\kappa
  W\frac{{\partial }K_{\rm hid}}{\partial z_j} +\frac{{\partial
  }W_{\rm hid}}{\partial z_j}\right) \left(\kappa W^{\dagger
  }\frac{{\partial }K_{\rm hid}}{\partial z_i^{\dagger }}
  +\frac{{\partial }W_{\rm hid}^{\dagger }}{\partial
  z_i^{\dagger}}\right) \nonumber \\ & & +{\rm e}^{\kappa
  K}\left(K^{-1}\right)^{\phi_a^{\dagger}\phi_b} \left(\kappa
  W\frac{{\partial }K_{\rm obs}}{\partial \phi_b} +\frac{{\partial
  }W_{\rm obs}}{\partial \phi_b}\right) \left(\kappa W^{\dagger
  }\frac{{\partial }K_{\rm obs}}{\partial \phi_a^{\dagger }}
  +\frac{{\partial }W_{\rm obs}^{\dagger }}{\partial
  \phi_a^{\dagger}}\right) \nonumber \\ & & -3\kappa {\rm e}^{\kappa
  K}WW^{\dagger} \nonumber \\ &\equiv & V_{1}+V_{2}+V_{3}-3\kappa {\rm
  e}^{\kappa K}WW^{\dagger} \, .
\end{eqnarray}
In the above expression, the inverse K\"ahlerian matrix is not the
total inverse matrix but the inverse matrix in each sector, \ie, for
instance, $\left(K^{-1}\right)^{d_{\alpha }^{\dagger}d_{\beta
}}=\left(K^{-1}_{\rm quint}\right)^{d_{\alpha }^{\dagger}d_{\beta
}}$. This is because these matrices involve derivatives of the
K\"ahler potential and, therefore, kill all the contributions which
are not in the sector considered. For convenience, we can separate
this potential into several parts, namely $V_1$, $V_2$ and $V_3$ as
defined above. The first term gives the potential
\begin{eqnarray}
V_{1} &=& {\rm e}^{\kappa K}\left(K^{-1}\right)^{d_{\alpha
}^{\dagger}d_{\beta }} \left(\kappa W_{\rm quint}\frac{{\partial
}K_{\rm quint}}{\partial d_{\beta }} +\frac{{\partial }W_{\rm
quint}}{\partial d_{\beta }}\right) \left(\kappa W^{\dagger }_{\rm
quint}\frac{{\partial }K_{\rm quint}}{\partial d_{\alpha }^{\dagger }}
+\frac{{\partial }W_{\rm quint}^{\dagger }}{\partial d_{\alpha
}^{\dagger}}\right)\nonumber \\ & & +{\rm e}^{\kappa
K}\left(K^{-1}\right)^{d_{\alpha }^{\dagger}d_{\beta }}\kappa
^2\frac{{\partial }K_{\rm quint}}{\partial d_{\beta }} \frac{{\partial
}K_{\rm quint}}{\partial d_{\alpha }^{\dagger}} \biggl[\left(W_{\rm
hid}+W_{\rm obs}\right)\left(W_{\rm hid}^{\dagger } +W_{\rm
obs}^{\dagger }\right) \nonumber \\ & & +W_{\rm quint}\left(W_{\rm
hid}^{\dagger }+W_{\rm obs}^{\dagger}\right) +W_{\rm quint}^{\dagger
}\left(W_{\rm hid}+W_{\rm obs}\right) \biggr] +{\rm e}^{\kappa
K}\left(K^{-1}\right)^{d_{\alpha }^{\dagger}d_{\beta }}\kappa
\nonumber \\ & & \times \left[ \frac{{\partial }K_{\rm
quint}}{\partial d_{\beta }} \frac{{\partial }W_{\rm quint}^{\dagger
}}{\partial d_{\alpha }^{\dagger }} \left(W_{\rm hid}+W_{\rm
obs}\right) +\frac{{\partial }K_{\rm quint}}{\partial d_{\alpha
}^{\dagger}} \frac{{\partial }W_{\rm quint}}{\partial d_{\beta }}
\left(W_{\rm hid}^{\dagger }+W_{\rm obs}^{\dagger }\right)\right]\, .
\end{eqnarray}
In the above formula, the first term usually gives rise to a
supergravity quintessence  potential while the other terms
represent the coupling of the quintessence field to the hidden and
observable sectors.

\par

Let us now consider the term $V_{2}$ associated with the hidden
sector. Using the fact that the K\"ahler potential is flat, it can
be written as
\begin{eqnarray}
V_{2} &=& {\rm e}^{\kappa K}\left(\kappa ^2W W^{\dagger
}z_iz_i^{\dagger }+\kappa W\frac{{\partial }W_{\rm hid}^{\dagger
}}{\partial z_i^{\dagger }}z_i^{\dagger } +\kappa W^{\dagger
}\frac{{\partial }W_{\rm hid}}{\partial z_i}z_i+\left\vert
\frac{{\partial }W_{\rm hid}}{\partial z_i}\right\vert ^2 \right)\, .
\end{eqnarray}
Finally, there is the observable matter part. Again, since the
K\"ahler potential is taken to be flat, the matrix
$\left(K^{-1}\right)^{\phi_a^{\dagger}\phi_b} $ is the unit
matrix. Therefore, one obtains
\begin{eqnarray}
\label{vobs}
V_{3} &=& {\rm e}^{\kappa K}\left(\kappa ^2W W^{\dagger
}\phi_a\phi_a^{\dagger }+\kappa W\frac{{\partial }W_{\rm obs}^{\dagger
}}{\partial \phi _a^{\dagger }}\phi_a^{\dagger } +\kappa W^{\dagger
}\frac{{\partial }W_{\rm obs}}{\partial \phi_a}\phi_a+\left\vert
\frac{{\partial }W_{\rm obs}}{\partial \phi_a}\right\vert ^2 \right)\,
.
\end{eqnarray}
So far, all these expressions are exact and only depend on the
assumption that we have three separate sectors in the theory. In the
following we will discuss the physical consequences of these
potentials and the coupling between the three sectors.

\subsection{Breaking Supersymmetry}

In the hidden sector, the supersymmetry breaking fields take a vacuum
expectation value (vev) obtained by solving the following equation
\begin{equation}
\frac{\partial V(z_j,Q,\phi_a)}{\partial z_i^\dagger}=0 \, .
\label{min}
\end{equation}
Solving the above expression for $\mean{z_i}$ leads to
\begin{equation}
\kappa ^{1/2}\mean{z_i}_{\rm min}=\kappa ^{1/2}\mean{z_i}_{\rm
min}\left(\mean{Q}, \mean{\phi _a}\right)\, .
\end{equation}
The previous considerations are valid at very high energies, well
above the electroweak transition. In this case, one has $\mean{\phi
_a}=0$. The presence of the quintessence sector implies that the
dynamics of the supersymmetry breaking sector is perturbed. This
cannot be neglected as the vev of the quintessence field is, {\it a
priori}, not negligible and leads to $Q$ dependent vev's, namely
$\kappa ^{1/2}\mean{z_i}_{\rm min}=\kappa ^{1/2}\mean{z_i}_{\rm
min}\left(Q,\mean{\phi _a}=0\right)$. As a consequence, if we
parameterize the hidden sector supersymmetry breaking in a model
independent way, we have
\begin{equation}
\label{parahidden}
\kappa ^{1/2}\mean{z_i}_{\rm min}\sim a_i(Q)\, , \quad \kappa
    \mean{W_{\rm hid}}_{\rm min}\sim M_{_{\rm S}}(Q)\, , \quad \kappa
    ^{1/2}\mean{\frac{\partial W_{\rm hid}}{\partial z_i}}_{\rm
    min}\sim c_i(Q)M_{_{\rm S}}(Q)\, ,
\end{equation}
where $a_i$ and $c_i$ are coefficients of order one which depend on
the detailed structure of the hidden sector. In the following, we
focus on the direction in field space where the hidden supersymmetry
breaking fields are located at their minimum satisfying
(\ref{min}). It is clear that, from the cosmological point of view,
having time varying fields $z_i$'s with vev's of the order of the
Planck mass can have drastic consequences. We will return to this
question below.

\par

The standard way to see whether supersymmetry is broken is to
calculate the $F$--terms of the theory that are defined, for the
fields in the hidden sector (similar expression would obviously hold
in the other sectors), by
\begin{equation}
F_{z_i}\equiv \left\langle {\rm e}^{\kappa K/2}\left( \kappa
W\frac{\partial K}{\partial z_i}+\frac{\partial W}{\partial
z_i}\right)\right\rangle \, .
\end{equation}
Explicit calculations lead to the following expression
\begin{eqnarray}
\label{defFz}
F_{z_i} &=& {\rm e}^{\kappa K_{\rm quint}/2+\sum _i\vert a_i\vert
^2/2} \frac{1}{\kappa ^{1/2}}\biggl[ \left(M_{_{\rm S}}+\kappa
  \left\langle W_{\rm
quint}\right \rangle \right)a_i+M_{_{\rm S}}c_i\biggr]\, .
\end{eqnarray}
In the observable sector, the corresponding result is obviously zero
but, in the quintessence sector, one generically obtains non vanishing
results.

\subsection{Computing the Soft Terms}

In this section, we still consider the theory at high energy, say at
the GUT scale. Our aim is then to calculate the soft
terms\cite{Brignole}, \ie the Lagrangian for the observable fields
after supersymmetry breaking. This means that in all the corresponding
expressions, we have to replace the hidden sectors terms by their
values at the minimum of the potential according to our general
parameterization given by Eqs.~(\ref{parahidden}). Let us also
introduce the gravitino mass. It is defined by
\begin{equation}
m_{3/2}\equiv \left\langle \kappa W {\rm e}^{\kappa K/2}\right\rangle
  \, .
\end{equation}
In the present context, the gravitino mass may depend on the
quintessence field and, therefore, may be a time-dependent quantity.
In the following, we write
\begin{equation}
\label{gravitino}
m_{3/2}={\rm e}^{\kappa K_{\rm quint}/2}m_{3/2}^0={\rm e}^{\kappa
K_{\rm quint}/2+\sum _i\vert a_i\vert ^2/2}\left(M_{_{\rm S}}+\kappa
\left\langle W_{\rm quint}\right \rangle \right)\, ,
\end{equation}
where $m_{3/2}^0$ is the mass that the gravitino would have without
the presence of the quintessence field (in this case, as already
discussed above, the coefficients $a_i$, $c_i$ and the mass $M_{_{\rm
S}}$ are constant and the term $W_{\rm quint }$ does not exist). The
soft terms are obtained by taking the limit $\mpl \rightarrow +\infty
$ in all the expressions while keeping the gravitino mass, hence the
mass $M_{_{\rm S}}$, fixed. This gives for the observable part of the
total potential
\begin{eqnarray}
V_{3} &=& {\rm e}^{\kappa K_{\rm quint}+\sum _i\vert
a_i\vert^2}\Biggl\{ \left[M_{_{\rm S}}^2+M_{_{\rm S}}\kappa
\left(W_{\rm quint}+W_{\rm quint}^{\dagger }\right)+\kappa ^2W_{\rm
quint}W_{\rm quint}^{\dagger }\right]\phi _a\phi _a^{\dagger} \nonumber
\\ & & +\left(M_{_{\rm S}}+\kappa W_{\rm quint}\right)\frac{{\partial
}W_{\rm obs}^{\dagger }}{\partial \phi_a^{\dagger}}\phi _a^{\dagger }
+\left(M_{_{\rm S}}+\kappa W_{\rm quint}^{\dagger }\right)
\frac{{\partial }W_{\rm obs}}{\partial \phi_a}\phi _a +
\left\vert \frac{{\partial }W_{\rm
obs}}{\partial \phi_a}\right\vert ^2 \Biggr\}\, .
\end{eqnarray}
Notice the terms $\exp(\sum _i \vert a_i\vert ^2) $ which remain
in the final expression of $V_{3}$. {\it A priori}, they depend on
the vev's of the quintessence field and introduce an additional
dependence on the quintessence field.

\par

Let us now consider the term $V_{2}$. Then, as before, one takes
the limit $\mpl \rightarrow +\infty $, keeping the gravitino mass
fixed and remembering that, {\it a priori}, $\mean{z_i}$ is of
order $\mpl$ (as it is the case for the quintessence field vev).
This gives
\begin{eqnarray}
\label{vhid}
V_{2} &=& {\rm e}^{\kappa K_{\rm quint}+\sum _i\vert a_i\vert ^2}
\Biggl\{ W_{\rm obs}\left[\left(M_{_{\rm S}}+\kappa W_{\rm
quint}^{\dagger }\right) \sum _i\vert a_i\vert^2+M_{_{\rm S}}\sum
_ia_ic_i\right]\nonumber \\ & & + W_{\rm
obs}^{\dagger}\left[\left(M_{_{\rm S}}+\kappa W_{\rm quint}\right)
\sum _i\vert a_i\vert^2+M_{_{\rm S}}\sum _ia_ic_i\right]
+\frac{1}{\kappa }\biggl(M_{_{\rm S}}^2+M_{_{\rm S}}\kappa W_{\rm
quint}+M_{_{\rm S}}\kappa W_{\rm quint}^{\dagger } \nonumber \\ & &
+\kappa ^2W_{\rm quint}W_{\rm quint}^{\dagger }\biggr)\sum _i \vert
a_i \vert ^2 +\frac{M_{_{\rm S}}}{\kappa }\biggl[\left(2M_{_{\rm
S}}+\kappa W_{\rm quint}+\kappa W_{\rm quint}^{\dagger }\right)\sum
_ia_ic_i \nonumber \\ & & +M_{_{\rm S}}\sum _i\vert c_i\vert
^2\biggr]\Biggr\}\, .
\label{vhid2}
\end{eqnarray}
The same expression can also be expressed as
\begin{eqnarray}
V_{2} &=& \sum _i\left\vert F_{z_i}\right\vert ^2 +{\rm e}^{\kappa
K_{\rm quint}+\sum _i\vert a_i\vert ^2} \Biggl\{ W_{\rm
obs}\left[\left(M_{_{\rm S}}+\kappa W_{\rm quint}^{\dagger }\right)
\sum _i\vert a_i\vert^2+M_{_{\rm S}}\sum _ia_ic_i\right]\nonumber \\ &
& + W_{\rm obs}^{\dagger}\left[\left(M_{_{\rm S}}+\kappa W_{\rm
quint}\right) \sum _i\vert a_i\vert^2+M_{_{\rm S}}\sum
_ia_ic_i\right]\Biggr\} \, ,
\end{eqnarray}
where we have used the expression of $F_{z_i}$, see also
Eq.~(\ref{defFz}). It is important to notice that the above potential
$V_2$ is made out of terms which are proportional to $\mpl ^0$ and to
$\mpl ^2$ (the terms proportional to negative powers of the Planck
mass having been eliminated by the limit $\mpl \rightarrow +\infty
$). On the contrary, $V_{\rm obs}$ is only made of terms proportional
to $\mpl ^0$. Therefore, one could worry that, in $V_{\rm hid}$, the
limit $\mpl \rightarrow +\infty $ is ill-defined. However, all the
terms proportional to $\mpl ^2$ appear in the term $\vert F_{z_i}\vert
^2$. The presence of this term is linked to the cosmological constant
problem and $\vert F_{z_i}\vert ^2$ will be fine-tuned in order to
avoid a large cosmological constant. For this reason, we do not need
to worry about these terms. On the other hand, the terms that
participate to the A and B terms in the soft SUSY breaking potential
have a perfectly well-defined limit $\mpl \rightarrow +\infty $.

\par

Finally, remains the term $V_{1}$. In the same limit, and taking
into account that the vev of $Q$ is {\it a priori} of the order of
the Planck mass, one obtains
\begin{eqnarray}
V_{1} &=& {\rm e}^{\kappa K}\left(K^{-1}\right)^{d_{\alpha
}^{\dagger}d_{\beta }} \left(\kappa W_{\rm quint}\frac{{\partial
}K_{\rm quint}}{\partial d_{\beta }} +\frac{{\partial }W_{\rm
quint}}{\partial d_{\beta }}\right) \left(\kappa W^{\dagger }_{\rm
quint}\frac{{\partial }K_{\rm quint}}{\partial d_{\alpha }^{\dagger }}
+\frac{{\partial }W_{\rm quint}^{\dagger }}{\partial d_{\alpha
}^{\dagger}}\right) \nonumber \\ & & +M_{_{\rm S}}^2 {\rm e}^{\kappa
K}\left(K^{-1}\right)^{d_{\alpha }^{\dagger}d_{\beta }}
\frac{{\partial }K_{\rm quint}}{\partial d_{\beta }} \frac{{\partial
}K_{\rm quint}}{\partial d_{\alpha }^{\dagger }} +M_{_{\rm S}} {\rm
e}^{\kappa K}\left(K^{-1}\right)^{d_{\alpha }^{\dagger}d_{\beta }}
\biggl[ \frac{{\partial }K_{\rm quint}}{\partial d_{\beta }}
\frac{{\partial }K_{\rm quint}}{\partial d_{\alpha }^{\dagger }}
\nonumber \\ & & \times \biggl(\kappa W_{\rm quint} +\kappa W_{\rm
quint}^{\dagger }\biggr) + \biggl( \frac{{\partial }K_{\rm
quint}}{\partial d_{\beta }} \frac{{\partial }W_{\rm quint}^{\dagger
}}{\partial d_{\alpha }^{\dagger }} + \frac{{\partial }K_{\rm
quint}}{\partial d_{\alpha }^{\dagger }} \frac{{\partial }W_{\rm
quint}}{\partial d_{\beta }}\biggr)\biggr] \nonumber \\ & & +W_{\rm
obs}{\rm e}^{\kappa K}\left(K^{-1}\right)^{d_{\alpha
}^{\dagger}d_{\beta }} \Biggl[\biggl(M_{_{\rm S}}+\kappa W_{\rm
quint}^{\dagger }\biggr)\kappa \frac{{\partial }K_{\rm
quint}}{\partial d_{\beta }} \frac{{\partial }K_{\rm quint}}{\partial
d_{\alpha }^{\dagger }} +\kappa \frac{{\partial }K_{\rm
quint}}{\partial d_{\alpha }^{\dagger }} \frac{{\partial }W_{\rm
quint}}{\partial d_{\beta }}\Biggr] \nonumber \\ & & +W_{\rm
obs}^{\dagger }{\rm e}^{\kappa K}\left(K^{-1}\right)^{d_{\alpha
}^{\dagger}d_{\beta }} \Biggl[\biggl(M_{_{\rm S}}+\kappa W_{\rm
quint}\biggr)\kappa \frac{{\partial }K_{\rm quint}}{\partial d_{\beta
}} \frac{{\partial }K_{\rm quint}}{\partial d_{\alpha }^{\dagger }}
+\kappa \frac{{\partial }K_{\rm quint}}{\partial d_{\beta}}
\frac{{\partial }W_{\rm quint}^{\dagger }}{\partial d_{\alpha
}^{\dagger} }\Biggr]\, .\nonumber \\
\end{eqnarray}
This part of the potential contains a coupling between the observable
superpotential and the quintessence field. It is of dimension $5$ in
the fields but is not suppressed by the Planck mass as $Q$ is of the
order of $\mpl$.

\par

Putting everything together, in particular taking into account the
term $-3WW^{\dagger }$, we find $V=V_{_{\rm DE}}+V_{_{\rm mSUGRA}}$
with
\begin{eqnarray}
\label{potde}
V_{_{\rm DE}} &=& {\rm e}^{\sum _i\vert a_i\vert ^2} V_{\rm quint} +
M_{_{\rm S}}^2 {\rm e}^{\kappa K_{\rm quint}+\sum _i\vert a_i\vert ^2}
\biggl[\left(K^{-1}\right)^{d_{\alpha }^{\dagger}d_{\beta }}
\frac{{\partial }K_{\rm quint}}{\partial d_{\beta }} \frac{{\partial
}K_{\rm quint}}{\partial d_{\alpha }^{\dagger }} -\frac{3}{\kappa
}\biggr] \nonumber \\ & & + M_{_{\rm S}} {\rm e}^{\kappa K_{\rm
quint}+\sum _i\vert a_i\vert ^2}
\Biggl\{\biggl[\left(K^{-1}\right)^{d_{\alpha }^{\dagger}d_{\beta }}
\frac{{\partial }K_{\rm quint}}{\partial d_{\beta }} \frac{{\partial
}K_{\rm quint}}{\partial d_{\alpha }^{\dagger }}-\frac{3}{\kappa
}\biggr] \biggl(\kappa W_{\rm quint} +\kappa W_{\rm quint}^{\dagger
}\biggr) \nonumber \\ & & + \left(K^{-1}\right)^{d_{\alpha
}^{\dagger}d_{\beta }}\biggl( \frac{{\partial }K_{\rm quint}}{\partial
d_{\beta }} \frac{{\partial }W_{\rm quint}^{\dagger }}{\partial
d_{\alpha }^{\dagger }} + \frac{{\partial }K_{\rm quint}}{\partial
d_{\alpha }^{\dagger }} \frac{{\partial }W_{\rm quint}}{\partial
d_{\beta }}\biggr)\Biggr\}+\sum _i\vert F_{z_i}\vert ^2\, ,
\end{eqnarray}
where $V_{\rm quint}$ is the potential that one would have obtained by
considering the dark sector alone, namely
\begin{eqnarray}
V_{\rm quint}(Q) &=& {\rm e}^{\kappa K_{\rm
quint}}\Biggl[\left(K^{-1}_{\rm quint}\right)^{d_{\alpha
}^{\dagger}d_{\beta }} \left(\kappa W_{\rm quint}\frac{{\partial
}K_{\rm quint}}{\partial d_{\beta }} +\frac{{\partial }W_{\rm
quint}}{\partial d_{\beta }}\right) \Biggl(\kappa W_{\rm
quint}^{\dagger }\frac{{\partial }K_{\rm quint}}{\partial d_{\alpha
}^{\dagger }} \nonumber \\ & & +\frac{{\partial }W_{\rm
quint}^{\dagger }}{\partial d_{\alpha }^{\dagger}}\Biggr)\nonumber
-3\kappa W_{\rm quint}W_{\rm quint}^{\dagger }\Biggr]\, .
\end{eqnarray}
Therefore, as expected, the breaking of supersymmetry has changed
the shape of the quintessence potential. As will be discussed in
the following, these corrections are important since they can in
principle modify the cosmological evolution of $Q$. On the other
hand, in the observable sector, one has
\begin{eqnarray}
\label{potobser}
V_{_{\rm mSUGRA}} &=& {\rm e}^{\kappa K_{\rm quint}+\sum _i\vert
a_i\vert^2}\left\vert \frac{{\partial }W_{\rm obs}}{\partial
\phi_a}\right\vert ^2 \nonumber \\ & & +{\rm e}^{\kappa K_{\rm
quint}+\sum _i\vert a_i\vert^2} \left[M_{_{\rm S}}^2+M_{_{\rm S}}
\left(\kappa W_{\rm quint}+\kappa W_{\rm quint}^{\dagger
}\right)+\kappa ^2W_{\rm quint}W_{\rm quint}^{\dagger }\right]\phi
_a\phi _a^{\dagger} \nonumber \\ & & + {\rm e}^{\kappa K_{\rm
quint}+\sum _i\vert a_i\vert^2}\biggl[ \left(M_{_{\rm S}}+\kappa
W_{\rm quint}\right)\frac{{\partial }W_{\rm obs}^{\dagger }}{\partial
\phi_a^{\dagger}}\phi _a^{\dagger } +\left(M_{_{\rm S}}+\kappa W_{\rm
quint}^{\dagger }\right) \frac{{\partial }W_{\rm obs}}{\partial
\phi_a}\phi _a \biggr] \nonumber \\ & & + {\rm e}^{\kappa K_{\rm
quint}+\sum _i\vert a_i\vert^2} W_{\rm obs}\Biggl\{\biggl(M_{_{\rm
S}}+\kappa W_{\rm quint}^{\dagger }\biggr)\biggl[\kappa
\left(K^{-1}\right)^{d_{\alpha }^{\dagger}d_{\beta }} \frac{{\partial
}K_{\rm quint}}{\partial d_{\beta }} \frac{{\partial }K_{\rm
quint}}{\partial d_{\alpha }^{\dagger }} \nonumber \\ & & +\sum
_i\vert a_i\vert ^2-3\biggr] +\kappa \left(K^{-1}\right)^{d_{\alpha
}^{\dagger}d_{\beta }} \frac{{\partial }K_{\rm quint}}{\partial
d_{\alpha}^{\dagger}} \frac{{\partial }W_{\rm quint}}{\partial
d_{\beta }}+M_{_{\rm S}}\sum _ia_ic_i\Biggr\} \nonumber \\ & & + {\rm
e}^{\kappa K_{\rm quint}+\sum _i\vert a_i\vert^2} W_{\rm obs}^{\dagger
}\Biggl\{\biggl(M_{_{\rm S}}+\kappa W_{\rm quint}\biggr)\biggl[\kappa
\left(K^{-1}\right)^{d_{\alpha }^{\dagger}d_{\beta }} \frac{{\partial
}K_{\rm quint}}{\partial d_{\beta }} \frac{{\partial }K_{\rm
quint}}{\partial d_{\alpha }^{\dagger }} \nonumber \\ & & +\sum
_i\vert a_i\vert ^2-3\biggr] +\kappa \left(K^{-1}\right)^{d_{\alpha
}^{\dagger}d_{\beta }} \frac{{\partial }K_{\rm quint}}{\partial
d_{\beta }} \frac{{\partial }W_{\rm quint}^{\dagger }}{\partial
d_{\alpha }^{\dagger }}+M_{_{\rm S}}\sum _ia_ic_i\Biggr\}\, .
\end{eqnarray}
This is the effective potential for the observable fields after
supersymmetry breaking. It contains a part involving the quintessence
field. It also contains the soft supersymmetry breaking terms of the
observable sector. Using the explicit form of the superpotential of
the mSUGRA model, see Eq.~(\ref{wobs}), the soft terms can easily be
estimated. The scalar potential in the observable sector takes the
form which precisely defines the soft terms
\begin{eqnarray}
V_{_{\rm mSUGRA}} &=&\cdots + {\rm e}^{\kappa K}V_{\rm
susy}+A_{abc}\left(\phi_a \phi _b\phi _c+\phi_a^{\dagger} \phi
_a^{\dagger }\phi _c^{\dagger }\right) +B_{ab}\left(\phi _a \phi
_b+\phi _a ^{\dagger }\phi _b^{\dagger }\right) \nonumber \\ & &
+m_{a\bar{b}}^2\phi _a
\phi _{b}^{\dagger }\, .
\end{eqnarray}
As a consequence, they read
\begin{eqnarray}
\label{softa} A_{abc} &=&
\lambda _{abc}{\rm e}^{\kappa K_{\rm
quint}+\sum _i\vert a_i\vert^2} \Biggl\{
\biggl(M_{_{\rm
S}}+\kappa W_{\rm quint}^{\dagger }\biggr)
+\frac13 \biggl(M_{_{\rm
S}}+\kappa W_{\rm quint}^{\dagger }\biggr)\biggl[\kappa
\left(K^{-1}\right)^{d_{\alpha }^{\dagger}d_{\beta }} \nonumber \\ & &
\times \frac{{\partial
}K_{\rm quint}}{\partial d_{\beta }} \frac{{\partial }K_{\rm
quint}}{\partial d_{\alpha }^{\dagger }} +\sum
_i\vert a_i\vert ^2-3\biggr] +\frac13 \kappa \left(K^{-1}\right)^{d_{\alpha
}^{\dagger}d_{\beta }} \frac{{\partial }K_{\rm quint}}{\partial
d_{\alpha}^{\dagger}} \frac{{\partial }W_{\rm quint}}{\partial
d_{\beta }}\nonumber \\ & & +\frac13 M_{_{\rm S}}\sum _ia_ic_i\Biggr\}\, ,\\
\label{softb} B_{ab} &=&
\mu _{ab}{\rm e}^{\kappa K_{\rm
quint}+\sum _i\vert a_i\vert^2} \Biggl\{
\biggl(M_{_{\rm
S}}+\kappa W_{\rm quint}^{\dagger }\biggr)
+\frac12 \biggl(M_{_{\rm
S}}+\kappa W_{\rm quint}^{\dagger }\biggr)\biggl[\kappa
\left(K^{-1}\right)^{d_{\alpha }^{\dagger}d_{\beta }} \nonumber \\ & &
\times \frac{{\partial
}K_{\rm quint}}{\partial d_{\beta }} \frac{{\partial }K_{\rm
quint}}{\partial d_{\alpha }^{\dagger }} +\sum
_i\vert a_i\vert ^2-3\biggr] +\frac12 \kappa \left(K^{-1}\right)^{d_{\alpha
}^{\dagger}d_{\beta }} \frac{{\partial }K_{\rm quint}}{\partial
d_{\alpha}^{\dagger}} \frac{{\partial }W_{\rm quint}}{\partial
d_{\beta }}\nonumber \\ & & +\frac12 M_{_{\rm S}}\sum _ia_ic_i\Biggr\}\, ,\\
\label{softm} m_{a\bar{b}} ^2 &=& {\rm e}^{\kappa K_{\rm
quint}+\sum _i\vert a_i\vert^2} \left[M_{_{\rm S}}^2+M_{_{\rm S}}
\left(\kappa W_{\rm quint}+\kappa W_{\rm quint}^{\dagger
}\right)+\kappa ^2W_{\rm quint}W_{\rm quint}^{\dagger }\right]\delta
_{a\bar{b}}\, .
\end{eqnarray}
This is the general form of the soft terms, calculated at the GUT
scale. Notice that as we have chosen the hidden and observable fields
to be canonically normalized, we find that the soft terms are
universal as expected. Here the soft terms are explicitly $Q$
dependent. This has severe consequences on the particle masses as we
find in the following section. Indeed, the previous result allows to
compute the $Q$-dependence of the fermions mass in the mSUGRA
model. As we discuss below, it is necessary to use the renormalization
group equations to be able to use the previous equations not only at
GUT scale but also at the electroweak transition.

\par

Finally, another soft term can be evaluated, the gaugino masses. As
discussed before, we assume that the gauge kinetic functions are given
by $f_{_{G}}=f_{_{G}}(d_\alpha,z_i)$. Then, we find that the
gaugino masses at GUT scale can be expressed as
\begin{equation}
\label{gauginomass}
\left(m_{1/2}\right)_{_{G}}=\frac{1}{f_{_{G}}+f_{_{G}}^{\dagger}}
\left(\sum_\alpha F^{d_\alpha} \frac{\partial f_{_{G}}}{\partial
d_\alpha} +\sum_i F^{z_i} \frac{\partial f_{_{G}}}{\partial
z_i}\right) \equiv {\rm e}^{\kappa K_{\rm
quint}/2}\left(m_{1/2}^0\right)_{_{G}}\, ,
\end{equation}
where the $F_{z_i}$ term has already been evaluated in
Eq.~(\ref{defFz}). Notice that it depends on an arbitrary function
$f_{_{G}}(d_\alpha,z_i)$ for each gauge group $G$.

\section{Electroweak Symmetry Breaking in Presence of Dark Energy}
\label{ESB}

\subsection{The Higgs Potential in Presence of Quintessence}

We now consider the application of the previous results to the
electroweak symmetry breaking since this is the way fermions in
the standard model are given a mass. We consider that the soft
terms have been run down from the GUT scale to the weak scale. The
next step consists in computing the potential in the Higgs sector
which belongs to the observable sector. In the MSSM, there are two
$\mbox{SU}(2)_{\rm L}$ Higgs doublets
\begin{equation}
H_{\rm u}=\begin{pmatrix} H_{\rm u}^+ \cr H_{\rm u}^0 \end{pmatrix} \,
, \quad H_{\rm d}=\begin{pmatrix} H_{\rm d}^0 \cr H_{\rm d}^-
\end{pmatrix}\, ,
\end{equation}
that have opposite hypercharges, \ie $Y_{\rm u}=1$ and $Y_{\rm
  d}=-1$. The only term which is relevant in the superpotential is
  $W_{\rm obs}=\mu H_{\rm u}\cdot H_{\rm d}+\cdots =\mu (H_{\rm
  u}^+H_{\rm d}^--H_{\rm u}^0H_{\rm d}^0)+\cdots $ such that the
  superpotential remains gauge invariant. For indices $a$, $b$ running
  in the Higgs sector, we have $\mu _{ab}=\mu \epsilon_{ab}$, where
  $\epsilon _{11}=\epsilon _{22}=0$ and $\epsilon
  _{12}=-\epsilon_{21}=1$. This term gives contribution to the
  globally susy term $V_{\rm susy}$, namely
\begin{equation}
V_{\rm susy}=\left \vert \mu \right \vert ^2\left(\left\vert H_{\rm
  u}^+\right\vert ^2+\left\vert H_{\rm u}^0\right\vert ^2+\left\vert
  H_{\rm d}^0\right\vert ^2+\left\vert H_{\rm d}^-\right\vert
  ^2\right)\, .
\end{equation}
Then, we have the contribution coming from the soft susy-breaking
terms. As can be seen in Eq.~(\ref{softb}), the coefficient $B_{ab}$
is proportional to $\mu _{ab}$ and we choose to write it as
$B_{ab}\equiv \mu B{\rm e}^{\kappa K_{\rm quint}}\epsilon _{ab}$,
where $B=B(Q)$ is now a function of the quintessence field, namely
\begin{eqnarray}
\label{bhiggs}
B(Q) &=& {\rm e}^{\sum _i\vert a_i\vert^2} \Biggl\{ \biggl(M_{_{\rm
S}}+\kappa W_{\rm quint}^{\dagger }\biggr) +\frac12 \biggl(M_{_{\rm
S}}+\kappa W_{\rm quint}^{\dagger }\biggr)\biggl[\kappa
\left(K^{-1}\right)^{d_{\alpha }^{\dagger}d_{\beta }} \nonumber \\ & &
\times \frac{{\partial }K_{\rm quint}}{\partial d_{\beta }}
\frac{{\partial }K_{\rm quint}}{\partial d_{\alpha }^{\dagger }} +\sum
_i\vert a_i\vert ^2-3\biggr] +\frac12 \kappa
\left(K^{-1}\right)^{d_{\alpha }^{\dagger}d_{\beta }} \frac{{\partial
}K_{\rm quint}}{\partial d_{\alpha}^{\dagger}} \frac{{\partial }W_{\rm
quint}}{\partial d_{\beta }}\nonumber \\ & & +\frac12 M_{_{\rm S}}\sum
_ia_ic_i\Biggr\}\, .
\end{eqnarray}
Then the B-soft susy-breaking term in the scalar potential can be
expressed as
\begin{equation}
V_{\rm B-soft}=\mu B(Q){\rm e}^{\kappa K_{\rm quint}}\left[H_{\rm
  u}^+H_{\rm d}^--H_{\rm u}^0H_{\rm d}^0+\left(H_{\rm
  u}^+\right)^{\dagger}\left(H_{\rm d}^-\right)^{\dagger
  }-\left(H_{\rm u}^0\right)^{\dagger }\left(H_{\rm
  d}^0\right)^{\dagger }\right]\, .
\end{equation}
Finally, the soft masses remain to be evaluated from
Eq.~(\ref{softm}). For this purpose, one writes
$m_{1\bar{1}}^2=m_{H_{\rm u}}^2{\rm e}^{\kappa K_{\rm quint}}$, and
$m_{2\bar{2}}^2=m_{H_{\rm d}}^2{\rm e}^{\kappa K_{\rm quint}}$, where
according to Eq.~(\ref{gravitino}) and~(\ref{softm}), $m_{H_{\rm
u}}=m_{H_{\rm d}}=m_{3/2}^0$ at the GUT scale. This degeneracy is
lifted by the renormalization group evolution as necessary to obtain
the radiative breaking of the electroweak symmetry~\cite{Savoy}, as
explained below. One obtains
\begin{equation}
V_{\rm m-soft}=m_{H_{\rm u}}^2{\rm e}^{\kappa K_{\rm quint}}
\left(\left\vert H_{\rm u}^+\right\vert ^2+\left\vert H_{\rm
u}^0\right\vert ^2\right) +m_{H_{\rm d}}^2{\rm e}^{\kappa K_{\rm
quint}} \left(\left\vert H_{\rm d}^0\right\vert ^2+\left\vert
H_{\rm d}^-\right\vert ^2\right)\, ,
\end{equation}
where according to the previous considerations, there is no reason to
assume that $m_{H_{\rm u}}=m_{H_{\rm d}}$, nor that $m_{H_{\rm
u}}^2>0$ at the electroweak scale. As is well-known the
renormalization group equations drive the initial value of $m_{H_{\rm
u}}^2=\left(m_{3/2}^0\right)^2$ towards negative values which is
necessary in order to have the electroweak transition.  This important
property is not modified by the presence of the quintessence
field. Let us notice that, in the above expressions, there are {\it a
priori} contributions coming from the K\"ahler potential in the Higgs
sector of the form ${\rm e}^{\kappa K_{\rm Higgs}}$. We ignore these
contributions since the vevs of the Higgs is very small in comparison
with $\mpl$.

\par

This is not yet the final expression of the potential because the
D-term remain to be calculated. They are given by
\begin{equation}
V_{_{\rm D}}^{_{\rm Higgs}}=\sum _{_{G}}\frac{g^2_{_G}}{2}\sum
_{\alpha }\left(G^a T^{\alpha }_{_G}\phi _a\right)\left(G^b T^{\alpha
}_{_G}\phi _b\right)\, ,
\end{equation}
where, in the summation, we have two gauge groups, namely
$\mbox{SU}(2)_{\rm L}$ (coupling constant $g$), $\mbox{U}(1)_{\rm Y}$
(coupling constant $g'$) and $T_{\alpha }$ are the generators of the
gauge groups under consideration. We recall that $G_a\equiv
K_a+\partial _aW/W$. Thanks to gauge invariance implying that
$\delta_G W=\partial _a W iT^\alpha_G \phi_a=0$, only the part
involving the K\"ahler potential gives a non vanishing
contribution. Then, a standard calculation leads to
\begin{equation}
V_{_{\rm D}}^{_{\rm Higgs}}=\frac18 \left(g^2+g'^2\right)
\left(\left\vert H_{\rm u}^+\right\vert ^2+\left\vert H_{\rm u}^0
\right\vert ^2-\left\vert H_{\rm d}^0\right\vert ^2-\left\vert H_{\rm
d}^-\right\vert ^2\right)^2+\frac12 g^2\left\vert H_{\rm u}^+H_{\rm
d}^{0\dagger}+H_{\rm u}^0H_{\rm d}^{-\dagger}\right\vert ^2\, .
\end{equation}
Notice that the D-term potential does depend on the quintessence field
due to the dependence of the GUT scale gauge coupling constants $g$
and $g'$ on the fields in the quintessence and hidden sector.

\par

The total Higgs potential is the sum of the F-term potential and of
the D-term. The explicit and complete expression reads
\begin{eqnarray}
V^{_{\rm Higgs}} &=& {\rm e}^{\kappa K_{\rm quint}}\Biggl\{
\left(\left \vert \mu \right \vert ^2{\rm e}^{\sum _i\vert a_i\vert
^2}+m_{H_{\rm u}}^2\right) \left(\left\vert H_{\rm u}^+\right\vert
^2+\left\vert H_{\rm u}^0\right\vert ^2\right)+ \left(\left\vert \mu
\right\vert ^2{\rm e}^{\sum _i\vert a_i\vert ^2}+m_{H_{\rm
d}}^2\right) \nonumber \\ & & \times \left(\left\vert H_{\rm
d}^0\right\vert ^2+\left\vert H_{\rm d}^-\right\vert ^2\right)+\mu
B(Q)\left[H_{\rm u}^+H_{\rm d}^--H_{\rm u}^0H_{\rm d}^0+\left(H_{\rm
u}^+\right)^{\dagger}\left(H_{\rm d}^-\right)^{\dagger }-\left(H_{\rm
u}^0\right)^{\dagger }\left(H_{\rm d}^0\right)^{\dagger
}\right]\Biggr\} \nonumber \\ & & +\frac18 \left(g^2+g'^2\right)
\left(\left\vert H_{\rm u}^+\right\vert ^2+\left\vert H_{\rm
u}^0\right\vert ^2-\left\vert H_{\rm d}^0\right\vert ^2-\left\vert
H_{\rm d}^-\right\vert ^2\right)^2+\frac12 g^2\left\vert H_{\rm
u}^+H_{\rm d}^{0\dagger}+H_{\rm u}^0H_{\rm d}^{-\dagger}\right\vert
^2\, .
\end{eqnarray}
As mentioned above, it contains an explicit $Q$ dependence from the
F--term potential but also from the D--term if the gauge functions
$f_{_{G}}$ are not trivial.

\subsection{Minimizing the Higgs potential}

We now investigate the minimum of this potential. Using a
$\mbox{SU}(2)_{\rm L}$ transformation, one can always require $H_{\rm
u}^+=0$. Then, writing $\partial V^{_{\rm Higgs}}/\partial H_{\rm
u}^+=0$, implies $H_{\rm d}^-=0$ as usual since the presence of the
quintessence dependent coefficient does not modify the structure of
the potential in terms of the Higgs fields. Moreover, $H_{\rm u}^0$
and $H_{\rm d}^0$ can be taken real since they have opposite
hypercharges. Finally, the potential reads
\begin{eqnarray}
\label{simpleVhiggs}
V^{_{\rm Higgs}}&=& {\rm e}^{\kappa K_{\rm quint}}\biggl[\left(\left
\vert \mu \right \vert ^2{\rm e}^{\sum _i\vert a_i\vert ^2}+m_{H_{\rm
u}}^2\right)\left\vert H_{\rm u}^0\right\vert ^2+ \left(\left \vert
\mu \right \vert ^2{\rm e}^{\sum _i\vert a_i\vert ^2}+m_{H_{\rm
d}}^2\right)\left\vert H_{\rm d}^0\right\vert ^2\nonumber \\ & & -2\mu
B(Q)\left\vert H_{\rm u}^0\right\vert \left\vert H_{\rm
d}^0\right\vert\biggr]+\frac18 \left(g^2+g'^2\right) \left( \left\vert
H_{\rm u}^0\right\vert ^2-\left\vert H_{\rm d}^0\right\vert
^2\right)^2 \, .
\end{eqnarray}
Given this potential, there are conditions for the existence of a
stable minimum for non vanishing values of the Higgs vevs. First, in
order to avoid that the potential be unbounded from below along the
direction $H_{\rm u}^0=H_{\rm d}^0$, we have to require that
\begin{equation}
\label{cond1}
2\left \vert \mu \right \vert ^2{\rm e}^{\sum _i\vert a_i\vert
^2}+m_{H_{\rm u}}^2+m_{H_{\rm d}}^2-2\mu B(Q) >0 \, .
\end{equation}
In addition, one must require that the origin be a saddle point which
leads to
\begin{equation}
\label{cond2}
\left(\left \vert \mu \right \vert ^2{\rm e}^{\sum _i\vert a_i\vert
^2}+m_{H_{\rm u}}^2\right)\left(\left \vert \mu \right \vert ^2{\rm
e}^{\sum _i\vert a_i\vert ^2}+m_{H_{\rm d}}^2\right)< \mu^2 B^2(Q)\, .
\end{equation}
If $\left \vert \mu \right \vert ^2{\rm e}^{\sum _i\vert a_i\vert
^2}+m_{H_{\rm u}}^2$ is negative, then the last condition is
automatically satisfied. As mentioned before, this is natural since
the renormalization group equations pushes towards negative values of
$m_{H_{\rm u}}^2<m_{H_{\rm d}}^2$. We now discuss this point in more
detail. Let us define the Higgs vevs as $\langle H_{\rm u}^0\rangle
\equiv v_{\rm u}$ and $\langle H_{\rm d}^0\rangle \equiv v_{\rm d}$.
The link between these vevs and the mass of the gauge bosons,
$W^{\pm}$, $Z^0$ is not modified by the presence of dark
energy. Technically, this is due to the fact that the kinetic terms of
the Higgs are standard because the corresponding K\"ahler potentials
are flat. The standard calculation leads to
\begin{equation}
\label{gmass}
m_{W^{\pm}}^2=\frac{g^2}{2}\left(v_{\rm u}^2+v_{\rm d}^2\right)\equiv
\frac{g^2}{2}v^2\, ,\quad
m_{Z^0}^2=\frac12\left(g^2+g'^2\right)\left(v_{\rm u}^2+v_{\rm
d}^2\right)\, .
\end{equation}
where the energy scale $v=v(Q)$ is equal to $\sim 174\mbox{GeV}$
today, \ie at vanishing redshift. The gauge boson masses pick up a
$Q$-dependence from the gauge coupling constants and from $v$.  For
the same reason as above, the cancellation of the photon mass leads to
$g'\cos \theta =g\sin \theta $, where $\theta $ is a $Q$--dependent
Weinberg angle via
\begin{equation}
\tan \theta \, (Q) = \frac{g'(Q)}{g(Q)}\, ,
\end{equation}
if the gauge functions $f_{_{G}}$ are non trivial. Despite the $Q$
dependence of $\theta $, we are guaranteed that the photon remains
massless. Finally, we define $\tan \beta \equiv v_{\rm u}/v_{\rm
d}$, or $v_{\rm u}=v \sin \beta $ and $v_{\rm d}=v\cos \beta $.
Then, the two loop expression for the renormalized Higgs masses
gives~\cite{Brax}
\begin{eqnarray}
\label{mhu}
m^2_{H_{\rm u}}\left( Q\right) &=& m^2_{H_{\rm d}}(Q)
-0.36\left(1+\frac{1}{\tan^2 \beta}\right)
\Biggl\{\left[m_{3/2}^0\left(Q\right)\right]^2\left(1-
\frac{1}{2\pi}\right)+8\left[m_{1/2}^0\left(Q\right)\right]^2
\nonumber \\ & & + \left(0.28 -\frac{0.72}{\tan ^2\beta }\right
)\left[A(Q) + 2m^0_{1/2}\right]^2\Biggr\}\, ,\\
\label{mhd}
m^2_{H_{\rm d}}\left(Q\right) &=&
      \left[m_{3/2}^0\left(Q\right)\right]^2\left(1-\frac{0.15}{4\pi}\right)
      + \frac{1}{2} \left[m^0_{1/2}\left(Q\right)\right]^2\, ,
\end{eqnarray}
where $m_{1/2}^0$ is the gaugino mass at GUT scale, see
Eq.~(\ref{gauginomass}). Let us notice that the gaugino mass is also a
$Q$-dependent quantity. In the previous expression, we have identified
$A_{abc}= {\rm e}^{\kappa K_{\rm quint}}A \lambda_{abc}$, see
Eq.~(\ref{softa}), which leads to
\begin{eqnarray}
A(Q) &=& {\rm e}^{\sum _i\vert a_i\vert^2} \Biggl\{ \biggl(M_{_{\rm
S}}+\kappa W_{\rm quint}^{\dagger }\biggr) +\frac13 \biggl(M_{_{\rm
S}}+\kappa W_{\rm quint}^{\dagger }\biggr)\biggl[\kappa
\left(K^{-1}\right)^{d_{\alpha }^{\dagger}d_{\beta }} \nonumber \\ & &
\times \frac{{\partial }K_{\rm quint}}{\partial d_{\beta }}
\frac{{\partial }K_{\rm quint}}{\partial d_{\alpha }^{\dagger }} +\sum
_i\vert a_i\vert ^2-3\biggr] +\frac13 \kappa
\left(K^{-1}\right)^{d_{\alpha }^{\dagger}d_{\beta }} \frac{{\partial
}K_{\rm quint}}{\partial d_{\alpha}^{\dagger}} \frac{{\partial }W_{\rm
quint}}{\partial d_{\beta }}\nonumber \\ & & +\frac13 M_{_{\rm S}}\sum
_ia_ic_i\Biggr\}\, .
\end{eqnarray}
The above expressions of the $Q$-dependent Higgs mass are obtained by
assuming than one starts at GUT scale with equal masses
($=m_{3/2}^0$), as indicated by the previous calculation of the soft
terms, see Eq.~(\ref{softm}). They are then evaluated at the
electroweak scale. After the electroweak breaking, the masses are
stabilized to the previous values. The choice of the parameters should
be made such that, in absence of any dark energy, the Higgs potential
has a minimum, \ie such that the conditions given by
Eqs.~(\ref{cond1}) and~(\ref{cond2}) are satisfied. Then, for
non-vanishing values of the quintessence field, it could very well
turn out that these conditions no longer hold.

\par

The next step is to perform the minimization of the Higgs potential
given by Eq.~(\ref{simpleVhiggs}). Straightforward calculations give
\begin{eqnarray}
\label{min1}
{\rm e}^{\kappa K_{\rm quint}}\left(\left \vert \mu \right \vert
^2{\rm e}^{\sum _i\vert a_i\vert ^2}+m_{H_{\rm u}}^2\right) &=& \mu
B(Q)\frac{{\rm e}^{\kappa K_{\rm quint}}}{\tan \beta
}+\frac{m_{Z^0}^2}{2}\cos \left(2\beta \right)\, , \\ 
\label{min2}
{\rm e}^{\kappa
K_{\rm quint}}\left(\left \vert \mu \right \vert ^2{\rm e}^{\sum _i
\vert a_i\vert ^2}+m_{H_{\rm d}}^2\right)&=&\mu B(Q){\rm e}^{\kappa
K_{\rm quint}}\tan \beta -\frac{m_{Z^0}^2}{2}\cos \left(2\beta
\right)\, .
\end{eqnarray}
Adding the two equations for the minimum, we obtain a quadratic  equation
determining $\tan \beta $. The solution can easily be found and
reads
\begin{eqnarray}
\label{tan}
\tan \beta (Q) &=& \frac{2\vert \mu \vert ^2{\rm e}^{\sum _i\vert
    a_i\vert ^2}+m_{H_{\rm u}}^2(Q)+m_{H_{\rm d}}^2(Q)}{2\mu
    B(Q)}\nonumber \\ & & \times \biggl(1 \pm \sqrt{1-4\mu
    ^2B^2(Q)\left[2\vert \mu \vert ^2{\rm e}^{\sum _i\vert a_i\vert
    ^2}+m_{H_{\rm u}}^2(Q)+m_{H_{\rm d}}^2(Q)\right]^{-2}}\biggr)\, .
\end{eqnarray}
{\it A priori}, this equation is a transcendental equation determining
$\tan \beta $ as $\tan \beta $ also appears in the right-hand-side of
the above formula, more precisely in the Higgs masses, see
Eqs.~(\ref{mhu}) and~(\ref{mhd}). However, if one performs an
expansion in $1/\tan \beta $, that is to say if $\beta $ is not too
far form $\pi/2$ then the factor $\tan \beta $ in Eqs.~(\ref{mhu})
and~(\ref{mhd}) can be forgotten and Eq.~(\ref{tan}) becomes an
algebraic expression giving the tangent of $\beta $. Another remark is
that we have in fact two solutions for $\tan \beta$, one corresponding
to the plus sign, the other corresponding to the minus sign.

\par

From the equations~(\ref{min1}) and (\ref{min2}), one can also deduce
how the scale $v$ depends on the quintessence field. This leads to
\begin{eqnarray}
v(Q)=\frac{2{\rm e}^{\kappa K_{\rm quint}/2}}{\sqrt{g^2+g'{}^2}}
\sqrt{\left\vert\left\vert \mu \right \vert ^2{\rm e}^{\sum _i\vert
    a_i\vert ^2}+m_{H_{\rm u}}^2\right\vert}+{\cal O}\left(\frac{1}{\tan
  \beta }\right)\, .
\end{eqnarray}
Notice the absolute value in the square root that insures the
positivity of the argument which is necessary since, as already
mentioned, the renormalization group equations pushes $m_{H_{\rm
u}}^2$ towards negative values. Through Eqs.~(\ref{gmass}), the above
equation can be used to calculate the gauge bosons mass explicitly.

\par

Then, finally, one has for the vevs of the two Higgs
\begin{eqnarray}
\label{vuvd}
v_{\rm u}(Q)&=&\frac{v(Q)\tan \beta (Q)}{\sqrt{1+\tan^2 \beta (Q)}}
=v(Q)+{\cal O}\left(\frac{1}{\tan ^2\beta }\right)\, ,\\ 
\label{vuvd2}
v_{\rm d}(Q) &=& \frac{v(Q)}{\sqrt{1+\tan ^2\beta (Q)}}
=\frac{v(Q)}{\tan \beta }+{\cal O}\left(\frac{1}{\tan ^2\beta
}\right)\, ,
\end{eqnarray}
at leading order in $1/\tan ^2\beta $ (but if we insert the expression
of $v$, then $v_{\rm u}$ and $v_{\rm d}$ are only determined at first
order in $1/\tan \beta $). It is necessary to perform the expansion in
$1/\tan ^2\beta $ in the above equations as it would be inconsistent
to keep higher order corrections in those formulae while neglecting
them in Eq.~(\ref{tan}). The two expressions~(\ref{vuvd})
and~(\ref{vuvd2}) constitute the main result of this article. It
implies that fermion masses become $Q$ dependent. Moreover there are
two kinds of masses, depending on whether the fermions couple to
$H_{\rm u}$ or $H_{\rm d}$
\begin{equation}
m_{{\rm u},a}^{_{\rm F}}(Q)= \lambda_{{\rm u},a}^{_{\rm F}} {\rm
  e}^{\kappa K_{\rm quint}/2+\sum _i\vert a_i\vert ^2/2}v_{\rm u}(Q)\,
  , \quad m_{{\rm d},a}^{_{\rm F}}(Q)=\lambda_{{\rm d},a}^{_{\rm F}}
  {\rm e}^{\kappa K_{\rm quint}/2+\sum _i\vert a_i\vert ^2/2}v_{\rm
  d}(Q)\, ,
\end{equation}
where $\lambda_{{\rm u},a}^{_{\rm F}}$ and $\lambda_{{\rm d},a}^{_{\rm
F}}$ are the Yukawa coupling of the particle $\phi_a$ coupling either
to $H_{\rm u}$ or $H_{\rm d}$. Clearly, in order to go further and
compute explicitly the $Q$-dependence of the fermions mass, it is
necessary to specify the hidden and quintessence sectors.

\par

Before discussing the above results, let us make the following
remark. Denoting by $v_{\rm u,d}(0)$ the vev in the absence of the
quintessence field we can introduce the coupling constants
\begin{equation}
A_{\rm u,d}(Q)\equiv {\rm e}^{\kappa K_{\rm quint}/2+\sum _i\vert
  a_i\vert ^2/2}\frac{v_{\rm u,d}(Q)}{v_{\rm u,d}(0)}\, .
\end{equation}
Using this coupling constant, the theory below the electroweak
scale becomes a scalar tensor theory
\begin{eqnarray}
S &=& -\frac{1}{2\kappa} \int {\rm d}^4 x \sqrt{-g} \left[R +\frac12
  g^{\mu \nu}\partial_{\mu} Q\partial _{\nu }Q+V_{_{\rm
  DE}}(Q)\right]+ S_{\rm mat}\left[\phi_{{\rm u},a}, A_{\rm u}^2(Q)
  g_{\mu\nu}\right] \nonumber \\ & & + S_{\rm mat}\left[\phi_{{\rm
  d},a}, A^2_{\rm d}(Q) g_{\mu\nu}\right]\, .
\end{eqnarray}
The quintessence field has not decoupled and corresponds to the scalar
sector of the theory. The action is written in the Einstein frame with
metric $g_{\mu\nu}$ while the matter action is split in two parts
corresponding to the matter fields $\phi^a_{\rm u,v}$ having coupled
either to $H_{\rm u}$ or $H_{\rm d}$. Notice that the presence of the
coupling to the metrics $A^2_{\rm u,v}(Q)g_{\mu\nu}$ implies that a
fifth force due to the scalar exchange of $Q$ is generated.  Moreover,
matter of type $\phi_{\rm u}$ couples differently to the metric than
$\phi_{\rm d}$. This is a violation of the weak equivalence principle.

\par

In addition, the presence of two types of matter with two couplings
$A_{\rm u,d}$ prevents one from defining a Jordan frame where the
gravitational constant depends on the quintessence field and matter
couples to the metric uniquely. This is very different from
scalar-tensor theories and string theory with a universal
dilaton. Even when string loops are taken into account, it has been
claimed in Ref.~\cite{DP} that a least coupling principle is at play
whereby the gravitational couplings to matter are proportional to a
single function of the dilaton. A cosmological attractor implies that
the dilaton runs towards the minimum of this function where
gravitational constraints are evaded. Here we have two intrinsically
different couplings. This is a feature of the MSSM coupled to
quintessence and comes from the necessity of having two Higgs fields
to cancel the gauge anomalies. Hence violations of the weak
equivalence principle are intrinsic when coupling quintessence to the
MSSM and/or mSUGRA.

\section{Discussion}
\label{Discussion}

The results obtained in the previous sections have three main
consequences: firstly, the shape of the quintessence potential is
modified as compared to what is obtained if one just considers a
separate dark energy sector, see Eq.~(\ref{potde}). Secondly, the mass
of the observable sector particles becomes $Q$-dependent
quantities. This is of course valid for the spin--$1/2$ fermions and
also for the gravitino and the gauge bosons. Thirdly, the gauge
couplings, among which is the fine structure constant, can also become
a function of the quintessence field depending on the choice of the
function $f_{_{G}}$. At this point, the following remark is in
order. The fact that the mass of the particles becomes $Q$-dependent
has already been noticed and some of the corresponding consequences
have been worked out, see for instance Ref.~\cite{Khoury}. However,
what is usually done is to keep the quintessence potential unchanged
and to postulate some dependence for the fermions masses (while, here,
we calculate this dependence from first principle). We see from the
previous considerations that this is not really consistent. In the
context of supergravity (and recall that the use of this framework is
mandatory since the vacuum expectation value of the quintessence is a
priori of the order of the Planck mass), particles acquire a mass
through the Higgs mechanism which, in turns, is possible only if
supersymmetry is broken, see the calculation of the soft terms in the
last section. But, as we have also seen, if supersymmetry is broken
then the quintessence potential is modified. This modification can of
course change the cosmological evolution of the quintessence field and
the determination of the free parameters characterizing the shape of
the potential (as, for instance, the typical mass scale).  Therefore,
when we take into account the fact that the fermions masses are
$Q$-dependent, it is also necessary to pay attention to the
corresponding modifications in the quintessence potential.

\subsection{Modifications of the Quintessence Potential}

Let us now discuss in more detail the three main consequences
described before. The new shape of the quintessence potential is only
known when the functions $a_i(Q)$ and $c_i(Q)$ are specified. {}From
Eq.~(\ref{potde}), we see that even if $a_i(Q)$ and $c_i(Q)$ vanish or
are just constant, that is to say if the vacuum expectation value of
the susy breaking fields $z_i$ are very small in comparison with the
Planck mass or if the $z_i$ fields are stabilized, the potential is
still not the original one. We can then envisage two different
situations. They are differentiated by the equation of state $w_Q$ of
the quintessence sector when $Q$ takes its present value in the
history of the universe. First of all, let us assume that the equation
of state is $w_Q\neq -1$.  This case can only be achieved when the
potential is of runaway type with an effective mass for the
quintessence field $m_Q\sim H_0$ the present Hubble rate. In this
case, this means that, despite the appearance of new terms in
$V_{_{\rm DE}}$, the original shape is stable and still presents a
runaway behavior.  Another possibility is that the new potential has a
minimum. In this situation, the field will have a completely different
mass in general and, from Eq.~(\ref{potde}), a typical expectation is
$m_Q\sim M_{_{\rm S}}\sim m_{3/2}^0$ although it should be possible to
avoid this conclusion in some particular cases. If the field is
stabilized at this minimum early in the history of the Universe, then
the effective equation of state is $w_Q=-1$ and the model is
equivalent to a cosmological constant. In this case, the quintessence
hypothesis clearly becomes useless.

\subsection{Fifth Force}

Let us now discuss the consequences of having $Q$-dependent masses. As
already mentioned before, this can lead to strong constraints coming
from gravitational experiments. Two situations can be envisaged
depending on the mass of the quintessence field.  If the mass of the
quintessence field is larger than $10^{-3} \mbox{eV}$, the
gravitational constraints are always satisfied as the range of the
force mediated by $Q$ is less than one millimeter. We see that this
typically occurs in the case where the quintessence potential has
developed a minimum, see above, first because, in such a situation,
the mass is expected to be $\sim m_{3/2}^0\gg 10^{-3}\mbox{eV}$ and
second because, if the field is stabilized at the minimum, then there
is no variation at all. Therefore, the case where there is a minimum
appears to be free from gravitational problems. On the other hand, we
have seen before that this is not a satisfactory model of dark
energy. If the mass is less than $10^{-3} \mbox{eV}$, the range of the
quintessence field is large and generically there will be violations
of the weak equivalence principle (as discussed in the next
subsection) and a large fifth force.  In order not to to be in
contradiction with fifth force experiments such as the recent Cassini
spacecraft experiment, one must require that the Eddington
(post-Newtonian) parameter $\vert \gamma -1\vert \le 5\times 10^{-5}$,
see Ref.~\cite{GR}. If one defines the parameter $\alpha _{\rm u,d}$
by
\begin{eqnarray}
\label{alpha}
\alpha_{\rm u,d}(Q) &\equiv & \left\vert \frac{1}{\kappa^{1/2}}
\frac{{\rm d}\ln m_{\rm u,d}^{_{\rm F}}(Q)}{{\rm d} Q}\right \vert
\nonumber \\
&=&\left \vert \frac{1}{\kappa^{1/2}} \frac{{\rm d}\ln \lambda _{\rm
u,d}^{_{\rm F}}(Q)}{{\rm d} Q}+\frac{1}{\kappa^{1/2}} \frac{{\rm d}\ln
\left[{\rm e}^{\kappa K_{\rm quint}/2+\sum _i\vert a_i\vert ^2/2}v_{\rm
u,d}(Q)\right]}{{\rm d} Q}\right \vert \, ,
\end{eqnarray}
then the difficulties are avoided by imposing that $\alpha_{\rm
u,d}^2\le 10^{-5}$ since one has $\gamma =1+\alpha _{\rm u,d}^2$.
This result is valid for a gedanken experiment involving the
gravitational effects on elementary particles. For macroscopic bodies,
the effects are more subtle and will be discussed later.  In this
case, in principle, one should precisely compute the $Q$ dependence of
$\alpha _{\rm u,d}$ to evaluate how serious the gravitational problems
are. Obviously, this cannot be done in detail unless the functions
$a_i(Q)$ and $c_i(Q)$ and the quintessence sector are known
exactly. However, one can give some generic arguments. Complying with
the Cassini bound probably requires either a fine--tuning of the
functions $a_i(Q)$ and $c_i(Q)$ such that the global $Q$-dependence of
the masses is almost canceled out. One could also use a non-minimal
setting whereby the Yukawa couplings $\lambda _{\rm u,d}^{_{\rm F}}$
become $Q$-dependent. In this last case, it could turn out that, in
Eq.~(\ref{alpha}), this dependence cancels exactly the dependence of
the Higgs vevs $v_{\rm u,d}(Q)$. This is a functional fine--tuning
which seems difficult to justify. If the Yukawa couplings are not
$Q$-dependent, then it is interesting to calculate the expressions of
the gravitational couplings $\alpha _{\rm u,d}$ using the results
obtained in the last section. One obtains
\begin{eqnarray}
\label{alphadefu} \alpha _{\rm u} &=& \frac{\kappa ^{1/2}}{2}\partial _Q 
K_{\rm quint}+\frac{\kappa ^{-1/2}}{2}\sum _i\partial _Q\vert a_i\vert
^2 +\frac{\kappa ^{-1/2}}{\tan \beta \left(1+\tan ^2\beta \right)}
\frac{{\rm d}\tan \beta }{{\rm d}Q}+\frac{\kappa ^{-1/2}}{v}\frac{{\rm
d}v}{{\rm d}Q}\, ,\\ 
\label{alphadefd}
\alpha _{\rm d}&=&\frac{\kappa ^{1/2}}{2}\partial _Q K_{\rm
quint}+\frac{\kappa ^{-1/2}}{2}\sum _i\partial _Q\vert a_i\vert ^2
-\frac{\kappa ^{-1/2}\tan \beta }{1+\tan ^2\beta } \frac{{\rm d}\tan
\beta }{{\rm d}Q}+\frac{\kappa ^{-1/2}}{v}\frac{{\rm d}v}{{\rm d}Q}\,
,
\end{eqnarray}
where the derivative of the function $\tan \beta (Q)$ can be expressed
as
\begin{eqnarray}
\label{dtan}
\frac{{\rm d}\tan \beta }{{\rm d}Q} &=& \left(\frac{{\rm d}m_{\rm
    H_{\rm u}}^2}{{\rm d}Q}+\frac{{\rm d}m_{\rm H_{\rm d}}^2}{{\rm
    d}Q}\right)\left(2\vert\mu \vert ^2 {\rm e}^{\sum _i\vert a_i\vert
    ^2} +m_{\rm H_{\rm u}}^2+m_{\rm H_{\rm d}}^2\right)^{-1}\tan \beta
    -\frac{1}{B(Q)}\frac{{\rm d}B(Q)}{{\rm d}Q}\tan \beta \nonumber \\
    & & \pm 2\mu \left(2\vert\mu \vert ^2 {\rm e}^{\sum _i\vert
    a_i\vert ^2} +m_{\rm H_{\rm u}}^2+m_{\rm H_{\rm d}}^2\right)^{-1}
    \Biggl[1-4\mu ^2B^2(Q)\biggl(2\vert\mu \vert ^2 {\rm e}^{\sum
    _i\vert a_i\vert ^2} +m_{\rm H_{\rm u}}^2 \nonumber \\ & & +m_{\rm
    H_{\rm d}}^2\biggr)^{-2}\Biggr]^{-1/2}\times \Biggl[-\frac{{\rm
    d}B(Q)}{{\rm d}Q} +B(Q)\left(\frac{{\rm d}m_{\rm H_{\rm
    u}}^2}{{\rm d}Q}+\frac{{\rm d}m_{\rm H_{\rm d}}^2}{{\rm
    d}Q}\right)\biggl(2\vert\mu \vert ^2 {\rm e}^{\sum _i\vert
    a_i\vert ^2} +m_{\rm H_{\rm u}}^2\nonumber \\ & & +m_{\rm H_{\rm
    d}}^2\biggr)^{-1}\Biggr]\, ,
\end{eqnarray}
assuming in this last equation, for simplicity, that the susy breaking
fields are stabilized at a vev small in comparison with the Planck
mass, that is to say $a_i=0$ but not necessarily $c_i=0$. The
derivatives of the soft terms $A$ and $B$ are known once the
quintessence sector is specified and one has from the renormalization
group equations~(\ref{mhu}) and (\ref{mhd})
\begin{eqnarray}
\frac{{\rm d}m_{\rm H_{\rm u}}^2}{{\rm d}Q} &\simeq &\frac{{\rm
d}m_{\rm H_{\rm d}}^2}{{\rm d}Q} -0.72\left(1-\frac{1}{2\pi
}\right)m_{3/2}^0\frac{{\rm d}m_{3/2}^0}{{\rm d}Q}-0.72\times
8m_{1/2}^0\frac{{\rm d}m_{1/2}^0}{{\rm d}Q} \nonumber \\ & &
-0.72\times 0.28 \left[\frac{{\rm d}A(Q)}{{\rm d}Q}+2\frac{{\rm
d}m_{1/2}^0}{{\rm d}Q}\right]\times \left[A(Q) + 2m_{1/2}^0\right]\, ,
\\ \frac{{\rm d}m_{\rm H_{\rm d}}^2}{{\rm d}Q} &\simeq &
2\left(1-\frac{0.15}{4\pi }\right)m_{3/2}^0\frac{{\rm
d}m_{3/2}^0}{{\rm d}Q} +m_{1/2}^0\frac{{\rm d}m_{1/2}^0}{{\rm d}Q}\, ,
\end{eqnarray}
the symbol ``approximate'' in the two last equations meaning that we
have used the fact that the terms in $1/\tan ^2\beta $ have been
neglected in the expression of the Higgs masses. In this case, one
should keep the leading order only in Eq.~(\ref{dtan}) which amounts
to
\begin{eqnarray}
\label{dtanTaylor}
\frac{{\rm d}\tan \beta }{{\rm d}Q} &\simeq & \left(\frac{{\rm d}m_{\rm
    H_{\rm u}}^2}{{\rm d}Q}+\frac{{\rm d}m_{\rm H_{\rm d}}^2}{{\rm
    d}Q}\right)\left(2\vert\mu \vert ^2 {\rm e}^{\sum _i\vert a_i\vert
    ^2} +m_{\rm H_{\rm u}}^2+m_{\rm H_{\rm d}}^2\right)^{-1}\tan \beta
    \nonumber \\ & &-\frac{1}{B(Q)}\frac{{\rm d}B(Q)}{{\rm d}Q}\tan
    \beta \, .
\end{eqnarray}
One should also evaluate the derivative of the scale
$v(Q)$. Straightforward manipulations lead to 
\begin{eqnarray}
\frac{1}{v}\frac{{\rm d}v}{{\rm d}Q} &\simeq & \frac{\kappa }{2}
\partial _QK_{\rm quint}-\frac{1}{g^2+g'{}^2}\left(g\frac{{\rm
d}g}{{\rm d}Q} +g'\frac{{\rm d}g'}{{\rm d}Q}\right)
+\frac{1}{2}\left(\left\vert \mu \right \vert ^2{\rm e}^{\sum _i\vert
a_i\vert ^2} +m_{\rm H_{\rm u}}^2\right)^{-1} \nonumber \\ & & \times
\left(\left\vert \mu \right\vert ^2{\rm e}^{\sum _i\vert a_i\vert ^2}
\sum _i\partial _Q\vert a_i\vert ^2+\frac{{\rm d}m_{\rm H_{\rm
u}}^2}{{\rm d}Q}\right)+{\cal O}\left(\frac{1}{\tan \beta }\right)\, .
\end{eqnarray}
In fact, in order to be fully consistent at leading order,
Eqs.~(\ref{alphadefu}) and (\ref{alphadefd}) should be written as
\begin{eqnarray}
\alpha _{\rm u} &=& \frac{\kappa ^{1/2}}{2}\partial _Q K_{\rm
quint}+\frac{\kappa ^{-1/2}}{2}\sum _i\partial _Q\vert a_i\vert ^2
+\frac{\kappa ^{-1/2}}{v}\frac{{\rm d}v}{{\rm d}Q}+{\cal
O}\left(\frac{1}{\tan ^2\beta }\right)\, ,\\ \alpha _{\rm
d}&=&\frac{\kappa ^{1/2}}{2}\partial _Q K_{\rm quint}+\frac{\kappa
^{-1/2}}{2}\sum _i\partial _Q\vert a_i\vert ^2 -\kappa
^{-1/2}\left(\frac{{\rm d}m_{\rm H_{\rm u}}^2}{{\rm d}Q}+\frac{{\rm
d}m_{\rm H_{\rm d}}^2}{{\rm d}Q}\right)\nonumber \\ & & \times
\left(2\vert\mu \vert ^2 {\rm e}^{\sum _i\vert a_i\vert ^2} +m_{\rm
H_{\rm u}}^2+m_{\rm H_{\rm d}}^2\right)^{-1} +\kappa ^{-1/2}\frac{{\rm
d}\ln B(Q)}{{\rm d}Q}+\frac{\kappa ^{-1/2}}{v}\frac{{\rm d}v}{{\rm
d}Q} \nonumber \\ & & +{\cal O}\left(\frac{1}{\tan ^2\beta }\right) \,
.
\end{eqnarray}
As already noticed above, if we use the explicit expression of the
derivative of $v(Q)$ evaluated before, then the above formulas giving
$\alpha _{\rm u}$ and $\alpha _{\rm d}$ are accurate to the order
$1/\tan \beta $ only. Let us also notice that, in principle, for a
given model, the transcendental equation~(\ref{tan}) can be solved
numerically and all the subsequent quantities evaluated by this method
if necessary.

\subsection{Violation of the Weak Equivalence Principle}

As already mentioned, we also have violations of the weak equivalence
principle. This is due to the fact that, in the MSSM/mSUGRA, the
fermions couple differently to the two Higgs doublets $H_{\rm u}$ and
$H_{\rm d}$. Technically, this can be seen as the consequence of
having two different scalar tensor theories with two different
conformal factor $A_{\rm u}(Q)$ and $A_{\rm d}(Q)$. Violations of the
weak equivalence principle are quantified in terms of the $\eta_{_{\rm
AB}}$ parameter defined by~\cite{DP,damour,JP}
\begin{equation}
\eta_{_{\rm AB}} \equiv \left(\frac{\Delta a}{a}\right)_{_{\rm AB}}
=2\frac{a_{_{\rm A}}-a_{_{\rm B}}}{a_{_{\rm A}}+a_{_{\rm B}}}\, ,
\end{equation}
for two test bodies A and B in the gravitational background of a third
one E.  Current limits~\cite{Su} indicate that $\eta_{_{\rm AB}}
=(+0.1\pm 2.7\pm 1.7)\times 10^{-13}$.

\par

The parameter $\eta _{_{\rm AB}}$ can be calculated in the following
way. The interaction between two bodies can be written as $-G(1+\alpha
_{_{\rm A}}\alpha _{_{\rm B}})m_{_{\rm A}}m_{_{\rm B}}/r_{_{\rm AB}}$,
where the dimensionless coefficient $\alpha _{_{\rm A}}$ is defined
according to Eq.~(\ref{alpha}), namely $\alpha _{_{\rm A}}=\kappa
^{-1/2}{\rm d}\ln m_{_{\rm A}}/{\rm d}Q$. Evaluating the parameter
$\eta _{_{\rm AB}}$ for $r_{_{\rm AE}}=r_{_{\rm BE}}$ leads to
\begin{equation}
\eta_{_{\rm AB}}\sim \frac{1}{2}\alpha_{_{\rm E}}\left(\alpha_{_{\rm
    A}}-\alpha_{_{\rm B}}\right)\, .
\end{equation}
For our gedanken experiment involving two elementary particles of
types u and d, this implies a strong violation of the equivalence
principle as $\eta_{\rm ud}$ can be large.

\par

For macroscopic bodies as the ones involved in the Cassini experiment
or the solar system tests, the gravitational effects involve the inner
structure of the tested bodies. In Ref.~\cite{DP}, it was shown that
the mass of an atom can be written as
\begin{equation}
\label{matom}
m_{\rm Atom}(Q)\sim \Lambda _{_{\rm QCD}}M+\sigma '(N+Z)+\delta '(N-Z)
+a_3\alpha _{_{\rm QED}}E\Lambda _{_{\rm QCD}}\, ,
\end{equation}
where $N$ is the number of neutrons and $Z$ the number of protons. A
priori, $\Lambda _{_{\rm QCD}}$ and $\alpha _{_{\rm QED}}$ are
$Q$-dependent quantities when the functions $f_{_{G}}$ are non
trivial. Their evolution is given by the running of the gauge group
couplings, namely
\begin{equation}
\label{runninggauge}
\frac{1}{\alpha_i(m)} = 4\pi f_i -\frac{b_i}{2\pi}\ln
\left(\frac{m_{_{\rm GUT}}}{m}\right)\, ,
\end{equation}
where $i=1$ corresponds to the group $G=\mbox{U}(1)_{\rm Y}$, $i=2$ to
$G=\mbox{SU}(2)$ and $i=3$ to $G=\mbox{SU}(3)$. In the above
expression, the gauge couplings are evaluated at a scale $m$ starting
from $m_{_{\rm GUT}}$, a high energy scale at which our effective
theory breaks down, chosen here to be the GUT scale. In the MSSM, the
coefficients $b_i$ are given by (for $i=1,\cdots, 3$)
$b_i=(-33/5,-1,3)$. We study the consequences of the $Q$--dependence
of the gauge couplings in the next subsection. Here, we just need to
recall that $\Lambda _{_{\rm QCD}}$ is the QCD scale $m$ at which
$\alpha _{_{\rm QCD}}$ blows up. Using Eq.~(\ref{runninggauge}), it
can be expressed as
\begin{equation}
\Lambda _{_{\rm QCD}}=m_{_{\rm GUT}}{\rm e}^{-8\pi ^2f_{3}/b_3}\, .
\end{equation}
In the previous expression, threshold corrections have been
neglected. They have been examined in more detail in
Refs.~\cite{dent,strass, calmet}. On the other hand, $\alpha _{_{\rm
QED}}$ is the fine structure constant and is linked to $\alpha _1$ and
$\alpha _2$ by
\begin{equation}
\alpha _{_{\rm QED}}(Q)=\frac{\alpha _2^2}{\alpha _1+\alpha _2}\, ,
\end{equation}
where the evolution of $\alpha _1$ and $\alpha _2$ can be obtained
from Eq.~(\ref{runninggauge}). Let us continue the description of the
quantities appearing in Eq.~(\ref{matom}). $M$ can be written as
$M=(N+Z)+E_{_{\rm QCD}}/\Lambda _{_{\rm QCD}}$, where $E_{_{\rm QCD}}$
is the strong interaction contribution to the binding energy of the
nucleus. The quantity $a_3\alpha _{_{\rm QED}}\Lambda _{_{\rm
QCD}}E=E_{_{\rm QED}}$, $E_{_{\rm QED}}$ being the Coulomb interaction
of the nucleus. $E$ is given by $E=Z(Z-1)/(N+Z)^{1/3}$. Finally, the
coefficients $\sigma '$ and $\delta '$ are given by
\begin{eqnarray}
\sigma ' &=& \frac12 \left(m_{\rm u}+m_{\rm d}\right)\left(b_{\rm
  u}+b_{\rm d}\right) +\frac{\alpha _{_{\rm QED}}}{2}\left(C_{\rm
  n}+C_{\rm p}\right)+\frac 12 m_{\rm e}\, , \\ \delta ' &=& -\frac12
  \left(m_{\rm u}-m_{\rm d}\right)\left(b_{\rm u}-b_{\rm d}\right)
  +\frac{\alpha _{_{\rm QED}}}{2}\left(C_{\rm n}-C_{\rm
  p}\right)-\frac 12 m_{\rm e}\, ,
\end{eqnarray}
where $b_{\rm u}$, $b_{\rm d}$, $C_{\rm n}/\Lambda _{_{\rm QCD}}$ and
$C_{\rm p}/\Lambda _{_{\rm QCD}}$ are dimensionless coefficients. The
masses $m_{\rm u,d}$ have already been defined before and $m_{\rm e}$
is the mass of the electron. From the above formulas, one sees that
the $Q$ dependence arises from the $Q$ dependences of $m_{\rm u,d}$,
$m_{\rm e}$ and the three functions $f_{i}$.

\par

It is then straightforward to calculate the coefficient $\alpha $ for
the atom A. It reads
\begin{eqnarray}
\kappa ^{1/2}\alpha_{_{\rm A}} &\sim &-\frac{8\pi^2}{b_3}
\frac{\partial f_{3}}{\partial Q} + \frac{N_{_{\rm A}}+Z_{_{\rm
A}}}{M_{_{\rm A}} }\frac{\partial }{\partial Q}\left(\frac{\sigma
'}{\Lambda _{_{\rm QCD}}}\right) +\frac{N_{_{\rm A}}-Z_{_{\rm
A}}}{M_{_{\rm A}} }\frac{\partial }{\partial Q}\left(\frac{\delta
'}{\Lambda _{_{\rm QCD}}}\right) \nonumber \\ & & +a_3 \frac{E_{_{\rm
A}}}{M_{_{\rm A}}} \frac{\partial \alpha _{_{\rm QED}}}{\partial Q}\,
,
\end{eqnarray}
where
\begin{eqnarray}
\frac{\partial }{\partial Q}\left(\frac{\sigma '}{\Lambda _{_{\rm
QCD}}}\right) &=& \frac12 \frac{\kappa ^{1/2}}{\Lambda _{_{\rm QCD}}}
\left(b_{\rm u}+b_{\rm d}\right)\left(\alpha _{\rm u}m_{\rm u}+\alpha _{\rm
d}m_{\rm d}\right) +\frac12 \frac{\kappa ^{1/2}}{\Lambda _{_{\rm QCD}}}\alpha
_{\rm d}m_{\rm e}+\frac{8\pi ^2}{b_3}\frac{\sigma '}{\Lambda _{_{\rm
QCD}}} \frac{\partial f_{3}}{\partial Q} \nonumber \\ & &
+\frac{C_{\rm n}+C_{\rm p}}{2\Lambda _{_{\rm QCD}}}
\frac{\partial \alpha _{_{\rm QED}}}{\partial Q}\, ,
\\
\frac{\partial }{\partial Q}\left(\frac{\delta '}{\Lambda _{_{\rm
QCD}}}\right) &=& -\frac12 \frac{\kappa ^{1/2}}{\Lambda _{_{\rm QCD}}}
\left(b_{\rm u}-b_{\rm d}\right)\left(\alpha _{\rm u}m_{\rm u}-\alpha
_{\rm d}m_{\rm d}\right) -\frac12 \frac{\kappa ^{1/2}}{\Lambda _{_{\rm
QCD}}}\alpha _{\rm d}m_{\rm e}+\frac{8\pi ^2}{b_3}\frac{\sigma
'}{\Lambda _{_{\rm QCD}}} \frac{\partial f_{3}}{\partial Q}
\nonumber \\ & & +\frac{C_{\rm n}-C_{\rm p}}{2\Lambda _{_{\rm
QCD}}}\frac{\partial \alpha _{_{\rm
QED}}}{\partial Q}\, ,
\end{eqnarray}
and
\begin{equation}
\frac{\partial \alpha _{_{\rm QED}}}{\partial Q}= -4\pi \frac{(2\alpha
  _1+\alpha _2)\alpha _2^3}{(\alpha _1+\alpha _2)^2}\frac{\partial
  f_{2}}{\partial Q} +4\pi \frac{\alpha _1^2\alpha _2^2}{(\alpha
  _1+\alpha _2)^2}\frac{\partial f_{1}}{\partial Q}\, .
\end{equation}
Let us notice that the coefficient $\alpha _{\rm d}$ appears in
front of the mass of the electron $m_{\rm e}$ because, in the
MSSM, the electron behaves as a ``d'' particle. For numerical
estimates of the previous expressions, one can use $\Lambda_{_{\rm
QCD}}\sim 180 \mbox{MeV}$, $b_{\rm u} + b_{\rm d}\sim 6$, $b_{\rm
u}-b_{\rm d}\sim 0.5$, $C_{\rm p} \alpha_{_{\rm QED}}\sim 0.63
\mbox{MeV}$, $C_{\rm n} \alpha_{_{\rm QED}}\sim -0.13 \mbox{MeV}$,
$\sigma'/\Lambda_{_{\rm QCD}}\sim 3.8\times 10^{-2}$,
$\delta'/\Lambda_{_{\rm QCD}}\sim 4.2 \times 10^{-4}$, $a_3
\alpha_{\rm QED} \sim 0.77 10^-3$, $m_{\rm u}\sim 5 \mbox{MeV}$,
$m_{\rm d}\sim 10 \mbox{MeV}$ and $m_{\rm e}\sim 0.5 \mbox{MeV}$.

\par

As discussed before, a conservative way of complying with the Cassini
results is to impose that $\vert \alpha_{_{\rm A}}\vert \le 10^{-3}$
in order to satisfy the constraint on the Eddington parameter. Knowing
the gravitational coupling we can express the violations of the weak
equivalence principle
\begin{eqnarray}
\eta_{_{\rm AB}} &=& \frac{1}{2}\kappa ^{-1/2}\alpha _{_{\rm E}}\biggl[
\frac{\partial }{\partial Q}\left(\frac{\sigma '}{\Lambda _{_{\rm
QCD}}}\right)\left(\frac{N_{_{\rm A}}+Z_{_{\rm A}}}{M_{_{\rm
A}}}-\frac{N_{_{\rm B}}+Z_{_{\rm B}}}{M_{_{\rm B}}}\right)\nonumber \\
& & +\frac{\partial }{\partial Q}\left(\frac{\delta '}{\Lambda _{_{\rm
QCD}}}\right)\left(\frac{N_{_{\rm A}}-Z_{_{\rm A}}}{M_{_{\rm A}}}
-\frac{N_{_{\rm B}}-Z_{_{\rm B}}}{M_{_{\rm B}}}\right)
+a_3\frac{\partial \alpha _{_{\rm QED}}}{\partial Q}\left(
\frac{E_{_{\rm A}}}{M_{_{\rm A}}} -\frac{E_{_{\rm B}}}{M_{_{\rm
B}}}\right)\biggr]\, .
\end{eqnarray}
In particular for two pairs of test bodies, the ratio $\eta_{_{\rm
AB}}/\eta_{_{\rm BC}}$ is independent of the background object E.

\par

Now although it is a tortuous route, the calculations of the fifth
force and equivalence principle violations are directly related to the
supergravity Lagrangian and can be computed from first principle.

\subsection{Variations of Constants}

Another consequence of the interaction between dark energy and the
observable sector is the variations of the gauge couplings. As
described before, this is linked to the choice of the functions
$f_{_{G}}(Q,z_i)$. A non-vanishing function $f_i$ implies, for
example, a variation of the gauge coupling constants,
\begin{equation}
\frac{\Delta f_i}{f_i}=-\frac{\Delta \alpha _i}{\alpha _i}\, ,
\end{equation}
evaluated at the grand unification scale $m_{_{\rm GUT}}$. To obtain
the low energy variations one must use the running of the fine
structure constant between the GUT scale and the weak scale of the
${\rm SU}(2)\times {\rm U}(1)_{\rm Y}$ coupling constants using
Eq.~(\ref{runninggauge}).

\par

Experimentally, the Webb {\it et al.} result~\cite{webb} suggests that
$\Delta \alpha_{_{\rm QED}}/\alpha_{_{\rm QED}}\sim (-0.76 \pm 0.28)
\times 10^{-5}$ since a redshift $z=3$. If confirmed, this is a strong
constraint on the $Q$ dependence of $f$. However, in contrast, other
groups~\cite{srianand} reported a null result from observations on the
Southern Hemisphere. One must also consider the Oklo bound $\vert
\Delta \alpha_{_{\rm QED}}/\alpha_{_{\rm QED}}\vert \le 10^{-7}$
between $z=0.5$ and now ~\cite{JP}.

\par

Another important constraint comes from the variation of the proton to
electron mass ratio $r\equiv m_{\rm p}/m_{\rm e}$ where
\begin{equation}
m_{\rm p}= C_{_{\rm QCD}}\Lambda _{_{\rm QCD}}+ b_{\rm u} m_{\rm u} +
b_{\rm d} m_{\rm d} + C_{\rm p} \alpha_{_{\rm QED}}\, ,
\end{equation}
where $C_{_{\rm QCD}}\sim 5.2$ is a constant. Using the results
obtained before, namely $m_{\rm e}= A_{\rm d}(Q)m_{\rm e}^0$, $m_{\rm
u,d}= A_{\rm u,d} (Q)m_{\rm u,d}^0$, we get
\begin{equation}
\frac{\Delta r}{r}\sim -\frac{8\pi^2}{b_3}\Delta f_3+b_{\rm
      u}\frac{m_{\rm u}}{m_{\rm p}}\Delta \alpha _{\rm u}
      +\left (b_{\rm d}\frac{m_{\rm d}}{m_{\rm p}}-1 \right )\Delta \alpha _{\rm d}
      +\frac{C_{\rm p}\alpha _{_{\rm QED}}}{m_{\rm p}} \frac{\Delta
      \alpha_{_{\rm QED}}}{\alpha_{_{\rm QED}}}\, .
\end{equation}
Experimentally this ratio was measured to be $\Delta r/r\sim ( 5.02\pm
1.81) \times 10^{-5}$ at $z=3$~\cite{petit} and, very recently, there
is also an indication for a possible variation, $\Delta r/r\sim (
2.0\pm 0.6) \times 10^{-5}$~\cite{reinhold}. If these results are
confirmed experimentally, this gives bounds on $\alpha_{\rm u,d}$ and
the variations of $f_i$.

\par

It should be noticed that, in the present framework, a variation of
the gauge couplings is not necessarily linked to a variation of the
particle masses. Variation of particle masses seems to be unavoidable
while variation of the gauge couplings depends on the choice of the
functions $f_i$. In particular, in a minimal setting, one could choose
a constant $f_i$ in which no variation of $\alpha _{_{\rm QED}}$ would
be present while the fermion masses would still be quintessence vev
dependent quantities.  However, a dependence on the hidden sector is
still necessary in order to generate gaugino masses. This may lead to
variations of the fine structure constant if the hidden sector fields
have $Q$-dependent vevs.

\subsection{Cold Dark and Baryonic Energy Densities}

Another consequence of having $Q$-dependent masses is that, {\it a
  priori}, the energy density of cold dark and baryonic matters no
  longer scales as $1/a^3$ but as
\begin{equation}
\rho \sim \frac{1}{a^3} \sum _a n_a m^{_{\rm F}}_{\rm u,d}\left(
\frac{Q}{\mpl}\right)\, ,
\end{equation}
where $n_a$ is the number of non-relativistic particles. The above
claim is justified by the fact that dark matter is usually considered
as belonging to the observable sector (for instance, the lightest
supersymmetric particle is often presented as the most popular
candidate for dark matter). It has been shown in Ref.~\cite{Khoury}
that this type of interaction between dark matter and dark energy can
result in an effective dark energy equation of state less than
$-1$. It is also interesting to notice that, even if the mass of the
quintessence field is larger than $10^{-3}\mbox{eV}$ in which case, as
explained before, all the gravitational tests are satisfied, there
still exists a signature, of cosmological nature, of the interaction
between dark matter and dark energy, at least as long as the
quintessence field is not stabilized at its minimum (if this one
exists). The coupling between matter and quintessence induces a
modification of the quintessence potential 
\begin{equation}
V_{_{\rm eff}}(Q)=V_{_{\rm DE}}(Q)+A_{_{\rm CDM}}(Q)\frac{\rho _{_{\rm
      CDM}}^0}{a^3}\, ,
\end{equation}
where $A_{_{\rm CDM}}(Q)\equiv m_{_{\rm CDM}}(Q)/m_{_{\rm
CDM}}(0)$. In the present context, $m_{_{\rm CDM}}$ is the mass of the
dark matter particle, typically the lightest supersymmetric
particle. In this case, it is necessary to diagonalize the neutralino
mass matrix and, as a result, the Q-dependence of $m_{_{\rm CDM}}$ can
be non trivial. When the effective potential admits a time-dependent
minimum, the model is known as a chameleon model~\cite{cham}.

\par

Another important constraint comes from the "kicks" received by
the quintessence field when particle species become
non-relativistic~\cite{DP,cham}. This may induce large variations
of the quintessence field. This is particularly dangerous for the
electron where the transition occurs during nucleosynthesis and
may lead to large mass variations threatening nucleosynthesis.

\section{Conclusions}
\label{Conclusion}

We have studied the coupling of quintessence to both observable and
hidden matter in supergravity. We have shown that the resulting
dynamics are modified by the breaking of supersymmetry in the hidden
sector. In particular, the shape of the quintessence potential is
changed and, consequently, its mass is no longer necessarily small
(\ie of the order of the Hubble parameter today). We have argued that,
if the potential acquires a minimum, then the mass becomes of the
order of the gravitino mass.

\par

We have also paid attention to the electroweak physics and the
influence of quintessence on the particle masses after electroweak
symmetry breaking. We have found that the particle masses become
quintessence dependent in a way which is parametrized by the dynamics
of the hidden sector and we have provided a general framework to
compute how the fermion masses depend on the quintessence field
vev. The resulting theory is equivalent to a scalar-tensor theory. If
the quintessence mass is greater than $10^{-3}\mbox{eV}$, which is
clearly the case when the mass of the quintessence field is of the
order of the gravitino mass, then the corresponding model is free from
gravitational problems. On the contrary, if $m_Q$ is smaller than
$10^{-3}\mbox{eV}$ then the range of the force is large and,
therefore, subject to the constraints existing on the presence of a
large fifth force and/or of a strong violation of the equivalence
principle.

\par

The main goal of the paper was to develop a general formalism which
allows us to compute $m(Q/\mpl)$ in Eq.~(\ref{yukcoupling}) in the
most general case, \ie without having to specify the details of the
underlying quintessence model. In a companion paper~\cite{BMcosmo},
we apply this formalism to concrete cases which illustrate the results
obtained here and investigate how the cosmological evolution of the
quintessence field is changed.

\par

For instance, the SUGRA model coupled to the observable and
supersymmetry breaking sectors receives drastic
modifications~\cite{BM1,BMcosmo}. Two alternatives are possible. In
the first one corresponding to stabilized (Q-independent) hidden
sector vev's, the quintessence potential develops a minimum at low
values of $Q$ where the mass of the quintessence field is of the order
of the gravitino mass. In this case gravitational problems are
evaded. Yet the cosmological evolution of the quintessence field is
such that it settles down at the minimum of the potential before Big
Bang Nucleosynthesis, implying that the model is indistinguishable
from a pure cosmological constant. On the other hand, when the hidden
sector vev's are not all stable, the potential can remain runaway. In
this case, constraints on the non-observation of violations of the
equivalence principle imply that $\kappa^{1/2} Q_{\rm now} \ll
1$. Scanning over the various models, as necessary to get concrete
predictions, requires a long discussion which is performed in detail
in Ref.~\cite{BMcosmo}.

\acknowledgments

We wish to thank Nicolas Chatillon, Gilles Esposito-Farese and Martin
Lemoine for enlightening discussions and careful reading of the
manuscript.


\clearpage \nocite{*}
\bibliography{quintpart}

\providecommand{\href}[2]{#2}\begingroup\raggedright\begin{thebibliography}{90}

\bibitem{LSS} Tegmark M {\it et al.}, Cosmological Parameters from
SDSS and WMAP, 2004 {\it Phys.~Rev.~D} {\bf 69} 103501
[astro-ph/0310723].

\bibitem{IA} Perlmutter S {\it et al.}, Measurements of Omega and
Lambda from 42 High-Redshift Supernovae, 1999 {\it Astrophys.~J.} {\bf
517} 565 [astro-ph/9812133]; Garnavich P M {\it et al.}, Constraints
on Cosmological from Hubble Space Telescope Observations of High-z
Supernovae, 1998 {\it Astrophys.~J.} {\bf 493} L53 [astro-ph/9710123];
Riess A G {\it et al.}, Observational Evidence from Supernovae for an
Accelerating Universe and Cosmological Constant, 1998 {\it Astron.~J.}
{\bf 116} 1009 [astro-ph/9805201]; Astier P {\it et al}, The supernova
Legacy Survey: Measurement of $\Omega_{\rm m}$, $\Omega_{\Lambda}$ and
$w$ from the First Year Data Set [astro-ph/0510447].

\bibitem{CMB} Spergel D N {\it et al.}, Wilkinson Microwave Anisotropy
Probe (WMAP) Three Years Results: Implications for Cosmology
[astro-ph/0603449]; Fosalba P, Gaztanaga E and Castander F, Detection
of the ISW and SZ Effecst from CMB-Galaxy Correlation, 2003 {\it
Astrophys.~J.} {\bf 597} L89 [astro-ph/0307249]; Scranton R {\it et
al.}, Physical Evidence of Dark Energy [astro-ph/0307335]; Boughn S
and Crittenden R, A Correlation of the Cosmic Microwave Sky with Large
Scale Structure, 2004 {\it Nature (London)} {\bf 427} 45
[astro-ph/0305001].

\bibitem{MSU} Martin J, Schimd C and Uzan J P, Testing for $w<-1$ in
  the Solar System, 2006 {\it Phys.~Rev.~Lett.} {\bf 96} 061303
  [astro-ph/0510208].

\bibitem{kachru} Kachru S, Schulz M and Silverstein E, Self-Tuning of
Flat Domain Walls in 5d Gravity and String Theory, 2000 {\it
Phys.~Rev.~D} {\bf 62} 045021 [hep-th/001206]; Arkani-Hamed N,
Dimopoulos S, Kaloper N and Sundrum R, A Small Cosmological Constant
for a Large Extra Dimensions, 2000 {\it Phys.~Lett.} {\bf B480} 193
[hep-th/000197].

\bibitem{deffayet} Deffayet C, Dvali G and Gavadadze G, Accelerated
Universe from Gravity Leaking to Extra Dimension, 2002 {\it
Phys.~Rev.~D} {\bf 65} 044023 [hep-th/0105068].

\bibitem{Lalak} Forste S, Lalak Z, Lavignac S and Nilles H P, A
Comment on Self--Tuning and vanishing Cosmological Constant in the
Brane World, 2000 {\it Phys.~Lett.} {\bf B481} 360 [hep-th/0002164].

\bibitem{Koyama} Gorbinov D, Koyama K and Sibiryakov S, More on Ghosts
in DGP Models [hep-th/0512097].

\bibitem{Susskind} Susskind L, The Anthropic Landscape of String
Theory [hep-th/0302219].

\bibitem{RP} Ratra B and Peebles P J E, Cosmological Consequences of a
  Rolling Homogeneous Scalar Field, 1988 {\it Phys.~Rev.~D} {\bf 37}
  3406.

\bibitem{quint} Wetterich C, Cosmologies with Variable Newton's
``Constant'', 1988 {\it Nucl.~Phys.} {\bf B302} 668; Wetterich C, The
Cosmon Model for an Asymptotically Vanishing Time-Dependent
Cosmological Constant, 1995 {\it Astron.~Astrophys.}  {\bf 301} 321
[hep-th/9408025]; Ferreira P G and Joyce M, Cosmology with a
Primordial Scaling Field, 1998 {\it Phys.~Rev.~D} {\bf 58}, 023503
[astro-ph/9711102].

\bibitem{PB} Bin\'etruy P, Models of Dynamical Supersymmetry Breaking
  and Quintessence, 1998 {\it Phys.~Rev.~D} {\bf 60} 063502
  [hep-ph/9810553]; Bin\'etruy P, Cosmological Constant Versus
  Quintessence, 2000 {\it Int.~J.~Theor.~Phys.}  {\bf 39}, 1859
  [hep-ph/0005037].

\bibitem{BM1} Brax P and Martin J, Quintessence and Supergravity, 1999
  {\it Phys.~Lett.} {\bf B468} 40 [astro-ph/9905040].

\bibitem{cope} Copeland E, Sami M and Tsujikawa S, Dynamics of Dark
  Energy, [hep-th/0603057].

\bibitem{BM2} Brax P and Martin J, The Robustness of Quintessence,
  2000 {\it Phys.~Rev.~D} {\bf 61} 103502 [astro-ph/9912046].

\bibitem{BMR1} Brax P, Martin J and Riazuelo A, Exhaustive Study of
Cosmic Microwave Background Anisotropies in Quintessential Scenarios,
2000 {\it Phys.~Rev.~D} {\bf 62} 103505 [astro-ph/0005428].

\bibitem{BMR2} Brax P, Martin J and Riazuelo A, Quintessence with Two
  Energy Scales, 2001 {\it Phys.~Rev.~D} {\bf 64} 083505
  [hep-ph/0104240].

\bibitem{Nilles} Nilles H P, Supersymmetry, Supergravity and Particle
  Physics, 1984 {\it Phys.~Rept.} {\bf 101} 1; Martin S P, A
  Supermmetry Primer, [hep-ph/9709356]; Aitchison I J R, Supersymmetry
  and the MSSM: An Elementary Introduction, Notes of Lectures for
  Graduate Students in Particle Physics, Oxford 1004 \& 2005.

\bibitem{GR} Will C M, The Confrontation between General Relativity
  and Experiment, 2006 {\it Living. Rev. Rel.} {\bf 9} 2
  [gr-qc/0510072]; Fischbach E and Talmadge C, The Search for
  non-Newtonian Gravity, 1999 {\it Springer-Verlag, New-York};
  Bertotti B, Iess L and Tortora P, A Test of General Relativity Using
  Radio Links with the Cassini Spacecraft, 2003 {\it Nature} {\bf 425}
  374; Esposito-Farese G [gr-qc/0409081].

\bibitem{Brignole} Brignole A, Ibanez L E and Munoz C, Soft Susy
Breaking Terms from Supergravity and String Theory, {\it Perspective
on Susy} Kane G editor [hep-ph/9707209].

\bibitem{Savoy} Derendinger J P and Savoy C, Quantum Effects and
  $SU(2)\times U(1)$ Breaking in Supergravity Gauge Theories, 1984
  {\it Nucl.~Phys.~B} {\bf 237} 307.

\bibitem{BMcosmo} Brax P and Martin J, The SUGRA Quintessence Model
Coupled to the MSSM, [astro-ph/0606306].

\bibitem{Brans} Brans C and Dicke R H, Mach Principle and a
  Relativistic Theory of Gravitation, 1961 {\it Phys.~Rev.~D} {\bf
  124} 925; Bergmann P G, Comments on the Scalar Tensor Theory, 1968
  {\it Int. J. Theor. Phys.} {\bf 1} 25; Nordtvedt K, PostNewtonian
  Metric for a General Class of Scalar Tensor Gravitational Theories
  and Observational Consequences, 1970 {\it Astrophys. J.} {\bf 161}
  1059; Wagoner R, Scalar Tensor Theory and Gravitational Waves, 1970
  {\it Phys. Rev. D} {\bf 1} 3216.

\bibitem{damour} Damour T, Testing the Equivalence Principle: Why and
How?, 1996 {\it Class. Quantum Grav.} {\bf 13} A33 [gr-qc/9606080]. 

\bibitem{amendola} Amendola L, Coupled Quintessence, 2000 {\it
  Phys.~Rev.~D} {\bf 62} 043511 [astro-ph/9908023].

\bibitem{FP} Farrar C R and Peebles P J E, Interacting Dark Matter and
  Dark Energy, 2004 {\it Astrophys. J.} {\bf 604} 1
  [astro-ph/0307316].

\bibitem{RS} Randall L and Sundrum R, Out of this World Supersymmetry
  Breaking, {\it Nucl. Phys. } {\bf 557} 79 [hep-th/9810155].

\bibitem{GV} Green M B and Vanhove P, Duality and Higher Derivatives
  in M Theory, {\it JHEP} {\bf 0601} 093 [hep-th/0510027].

\bibitem{Mu}
Armendariz-Picon C, Mukhanov V and Steinhardt P, Essentials of
K-essence, {\it Phys.~Rev.~D} {\bf 63} 103510 [astro-ph/0006373]. 

\bibitem{Brax} Brax P and Savoy C, Models with Inverse Sfermion Mass
 Hierarchy and Decoupling of the SUSY FCNC Effects, 2000 {\it JHEP}
 {\bf 0007} 048 [hep-ph/0004133].

\bibitem{DP} Damour T and Polyakov A M, The String Dilaton and the
Least Coupling Principle, 1994 {\it Nucl. Phys. } {\bf B423} 532
[hep-th/9401069].

\bibitem{Khoury} Das S, Corasiniti P S and Khoury J,
  Super-acceleration as Signature of Dark Sector Interaction,
  [astro-ph/0510628].

\bibitem{JP} Uzan J P, The Fundamental Constants and their Variations:
  Observational Status and Theoretical Motivations, 2003 {\it
  Rev. Mod. Phys.} {\bf 75} 403 [hep-ph/0205340].

\bibitem{Su} Su Y {\it et al}, New Tests of the Universality of Free
  Fall, 1994 {\it Phys.~Rev.~D} {\bf 50} 3614; Baessler {\it et al},
  Improved Test of the Equivalence Principle for Gravitational
  Self-Energy, 1999 {\it Phys. Rev. Lett.} {\bf 83} 3585; Adelberger E
  G, New Tests of Eisntein's Equivalence Principle and Newton's
  Inverse-Squared Law, 2001 {\it Class. Quantum Grav.} {\bf 18} 2397;
  Williams J G, Turyshev S G and Boggs D H, Progress in Lunar Ranging
  Tests of Relativistic Gravity, 2004 {\it Phys. Rev. Lett.} {\bf 93}
  261101 [gr-qc/0411113].

\bibitem{dent} Dent T and Fairbairn M, Time-varying Coupling
  Strengths, Nuclear Forces and Unification, 2003 {\it Nucl. Phys. B}
  {\bf 653} 256 [hep-ph/0112279].

\bibitem{strass} Langacker P, Segre G and Strassler M, Implications of 
Gauge Unification for Time Variation of the Fine Structure Constant,
2002 {\it Phys. Lett.} {\bf B528} 121 [hep-ph/0112233].

\bibitem{calmet} Calmet X and Fritzsch H, Symmetry Breaking and Time
Variation of Gauge Couplings, 2002 {\it Phys. Lett.} {\bf B540} 173
[hep-ph/0204258].

\bibitem{webb} Webb J K {\it et al}, Further Evidence for Cosmological
  Evolution of the Fine Structure Constant, {\it Phys. Rev. Lett.}
  {\bf 87} 091301 (2001) [astro-ph/0012539].

\bibitem{srianand} Srianand R, Chand H, Petitjean P and Aracil B,
  Limits on the Time Variation of the Electromagnetic Fine-Structure
  Constant in the Low Energy Limit from Absorption Lines in the
  Spectra of Distant Quasars, 2004 {\it Phys. Rev. Lett.} {\bf 92}
  121302; Quast R, Reimers D and Levshakov S, Probing the Variability
  of the Fine-Structure Constant with the VLT/UVES, 2004 {\it
  Astron. Astrophys.} {\bf 415} L7 [astro-ph/0311280].

\bibitem{petit} Ivanchik A {\it et al}, Does the Photon-to-Electron
  Mass Ratio Vary in the Course of the Cosmological Evolution?  (2003)
  {\it Astrophys. Space. Sci.} {\bf 283} 583 (2003)
  [astro-ph/0210299].

\bibitem{reinhold} Reinhold E {\it et al}, Indication of a
  Cosmological Variation of the Proton-Electron Mass Ratio Based on
  Laboratory Measurement and Reanalysis of $\mbox{H}_2$ Spectra, 2006
  {\it Phys. Rev. Lett.}  {\bf 96} 151101.

\bibitem{cham} Brax P, van de Bruck C, Davis A C, Khoury J and Weltman
  A, Detecting Dark Energy in Orbit--the Cosmological Chameleon,
  (2004) {\it Phys. Rev. D } {\bf 70} 123518 [astro-ph/0408415].

\end{thebibliography}\endgroup
\bibliographystyle{plain}

\end{document}